\documentclass[letterpaper,twocolumn,10pt]{article}
\usepackage{usenix-2020-09}

\usepackage{tikz}

\usepackage{amsmath}
\usepackage{soul}
\usepackage{xspace}
\usepackage{hyperref}
\usepackage{xurl}
\usepackage[normalem]{ulem}
\usepackage{enumitem}
\usepackage{booktabs}
\usepackage{graphicx}
\usepackage{caption}
\usepackage{comment}
\usepackage{subcaption}
\usepackage{epstopdf}
\usepackage[]{authblk}

\usepackage{titlesec}

\setlist[itemize]{leftmargin=*}
\setlist[description]{leftmargin=*}

\newcommand{\yang}[1]{\textbf{\color{blue}[Yang: #1]}}

\newcommand{\sys}[0]{FT-HSDP\xspace}

\begin{document}

\date{}

\title{\Large Training LLMs with Fault Tolerant HSDP on 100,000 GPUs}

\author[$1$]{Omkar Salpekar}
\author[$1$]{Rohan Varma}
\author[$2$]{Kenny Yu}
\author[$1$]{Vladimir Ivanov}
\author[$1$,$3$]{Yang Wang}
\author[$1$]{Ahmed Sharif}
\author[$1$]{Min Si}
\author[$1$]{Shawn Xu}
\author[$1$]{Feng Tian}
\author[$1$]{Shengbao Zheng}
\author[$1$]{Tristan Rice}
\author[$1$]{Ankush Garg}
\author[$1$]{Shangfu Peng}
\author[$1$]{\\Shreyas Siravara}
\author[$1$]{Wenyin Fu}
\author[$1$]{Rodrigo de Castro}
\author[$1$]{Adithya Gangidi}
\author[$1$]{Andrey Obraztsov}
\author[$1$]{\\Sharan Narang}
\author[$4$]{Sergey Edunov}
\author[$1$]{Maxim Naumov}
\author[$1$]{Chunqiang Tang}
\author[$1$]{Mathew Oldham}

\affil[$1$~]{Meta Platforms}
\affil[$2$~]{Thinking Machines Labs}
\affil[$3$~]{The Ohio State University}
\affil[$4$~]{Genesis Molecular AI}

\maketitle


\begin{abstract}

Large-scale training systems typically use synchronous training, requiring all GPUs to be healthy simultaneously. In our experience training on O(100K) GPUs, synchronous training results in a low efficiency due to frequent failures and long recovery time.

To address this problem, we propose a novel training paradigm, Fault Tolerant Hybrid-Shared Data Parallelism (\sys). \sys uses data parallel replicas as units of fault tolerance. When failures occur, only a single data-parallel replica containing the failed GPU or server is taken offline and restarted, while the other replicas continue training.
To realize this idea at scale, \sys incorporates several techniques:
1) We introduce a Fault Tolerant All Reduce (FTAR) protocol for gradient exchange across data parallel replicas. 
FTAR relies on the CPU to drive the complex control logic for tasks like adding or removing participants dynamically,
and relies on GPU to perform data transfer for best performance.
2) We introduce a non-blocking catch-up protocol, allowing a recovering replica to join training with minimal stall.

Compared with fully synchronous training at O(100K) GPUs, \sys can reduce the stall time due to failure recovery 
from 10 minutes to 3 minutes, increasing effective training time from 44\% to 80\%. 
We further demonstrate that \sys's asynchronous recovery does not bring
any meaning degradation to the accuracy of the result model.

\end{abstract}

\section{Introduction} 

Training large transformer-based models with tens or hundreds of thousands of accelerators has become a research and engineering challenge of significant interest in recent years. For example, Llama 3 models were scaled up to 405 Billion parameters across 16K GPUs~\cite{llama3,llama3-parallelism}. Subsequent models sought to further increase the total parameter count, increasing
training size to a total of 100K - 200K GPUs~\cite{llama4,semianalysis,200kGPUs}.

Most large-scale training systems use fully synchronous training, where all GPUs must be simultaneously healthy in order for the training to make progress.
If a GPU or server fails, all members must restart from the last checkpoint~\cite{RobustLLM}.
In addition, collective libraries typically require all members to be known on initialization, and reconfiguring members dynamically is  not supported.
Therefore, all members must also reinitialize communication connections, which can cause a long stall at a large scale (Section~\ref{sec:motivation}).

For scale larger than 32K GPUs, a synchronous training approach faces multiple scalability challenges.
First, as the scale grows, the failure rate also increases. According to estimates from our studies, with 100K GPUs, we will experience a failure every 18 minutes. 
Second, the stalling time of synchronous recovery also increases. Even after devoting a significant amount of engineering effort to reducing
restart times, our experiments show that synchronous recovery takes up to 10 minutes at 100K GPUs.
This means that, out of these 18 minutes, 10 minutes will be spent on failing over and restarting. As a result, there are only 8 minutes for effective training. This results in an effective training time of $\frac{8}{18}$ = 44\%, which is unacceptable.

To reduce stalling due to recovery at a large scale, we introduce Fault Tolerant Hybrid-Shared Data Parallelism (\sys). Following
the idea of HSDP~\cite{hsdp-aws,fsdp}, \sys creates multiple replicas. Each replica is responsible
for training a subset of input data and consists of thousands of GPUs organized using a mix of parallelism
techniques such as  data, tensor, pipeline, expert, and context parallelism~\cite{fsdp,shoeybi2019megatron,controlled_memory_zero_bubble_pp,deepseekai2025deepseekv3technicalreport,dao2023flashattention,shah2024flashattention}.
GPUs from different replicas
periodically exchange gradients in a data parallelism manner.
Replication presents two opportunities for fault tolerance. First, if a failure occurs, \sys only needs to rebuild
the replica that includes the failed node, which reduces the
scale of recovery and thus reduces recovery time. Second, replication enables asynchronous recovery:
While the failed replica is recovering,
healthy replicas can still proceed with training.

Although this idea has been explored in previous work~\cite{lin2019dynamicminibatchsgdelastic,Gandhi2024ReCycle},
two important questions remain: 1) How to realize this idea at scale? 2) Does asynchronous recovery have any impact
on numerical accuracy and model quality?

\vspace{-.1in}
\paragraph{Recovery at scale with minimal stall.} We meet several challenges when
realizing asynchronous recovery at scale.

First, to support failure recovery, a node needs to be able to
detect failed or recovering nodes and remove or add connections dynamically.
Implementing such logic is straightforward on CPUs but is challenging 
in the GPU-driven NCCL communication~\cite{nccl}, which is widely used today due to its superior performance.
The fundamental problem is that, as a device designed mainly
for massive parallel communication, GPU still lacks capabilities
to implement complicated control logic, such as treating different types
of failure differently, removing or establishing connections dynamically, congestion control, etc.
As a result, our prior prototype and others~\cite{sergeev2018horovodfasteasydistributed,RobustLLM} need to reboot all nodes
to reconstruct connections, which can take several minutes and largely negates
the benefits of asynchronous recovery.
Using HSDP replicas as the units of fault tolerance simplifies reconfiguration,
since there is only one collective (all reduce) that spans multiple replicas.
Leveraging this observation,
we have built the Fault Tolerance All Reduce (FTAR) protocol, which relies on the CPU to drive
the complex control logic and relies on the GPU to perform the data transfer.
Our experiments show that FTAR can achieve a performance
comparable to that of NCCL.

Second, when a recovering replica re-joins \sys, it needs to ``catch up'' with
other replicas by loading the latest checkpoint. 
However, when the recovering replica is
fetching the checkpoint, other replicas are training new data, creating
a gap between the recovering replica and other replicas.
In a traditional replicated system, mitigating the gap
typically requires stalling or slowing down the other replicas~\cite{Wang2012Gnothi,Shi2016Cheap}. To minimize such stalling,
\sys incorporates two techniques. First, it introduces
a non-blocking catch up protocol by leveraging the specific properties
of training: Assuming that the recovering replica fetches and loads checkpoint in step n,
and healthy replicas train new data in step n, the recovering replica can send a zero
gradient at the end of step n and then the gradient exchange process can 
bring all replicas to the same state. Second, \sys fully overlaps the fetching of checkpoints on the
recovering replica with the training on healthy replicas. This means that the recovering replica, which may consist of thousands of GPUs, must finish fetching checkpoints
within tens of seconds. To achieve this, \sys has incorporated
a peer-to-peer checkpoint fetching protocol, which allows recovering GPUs
to fetch checkpoints from other GPUs directly in a load-balanced manner.

We measure the recovery stalling time of \sys in a real setting with 98K GPUs.
Our evaluation shows that, compared to the fully synchronous training recovery approach,
which incurs 10 minutes of stalling, \sys can reduce the stalling time to 3 minutes.
This increases the effective training time from 44\% to 80\%.

\vspace{-.1in}
\paragraph{Numerical accuracy.} Due to asynchronous recovery, \sys trains a varying amount of data per step. There
is a general concern about whether such approaches will affect the accuracy or even the convergence
of training. To answer this question,
we test how failure recovery in \sys can affect training accuracy
by extrapolating from experiment results in a smaller setting with 256 GPUs. Our results show
that \sys recovery
does not have a significant impact on final model quality, although it incurs some variance during training.
We further show that square root learning rate intervention can flatten such variance.

\vspace{-.1in}
\paragraph{Contributions.}
This paper makes the following contributions:
1. We provide a detailed breakdown of time spent in different steps of synchronous recovery
at the scale of O(100K) GPUs, which demonstrates the unfeasibility at such scales.
2. We present a fault tolerant all reduce protocol for fast collective communication in the failure-free case and flexible control in the failure case.
3. We present a non-blocking recovery protocol that minimizes stalled time when replicas recover.
4. We show that \sys's asynchronous recovery does not bring a significant impact on model accuracy.

\section{Background}\label{sec:background}

The specific training scenario we consider is training a large (in the range of 70B-900B active parameters) language model on O(trillions) of tokens across a 100K GPU training cluster. 

\subsection{Networking}
\label{sec:networking}

A training cluster with over 100K GPUs inevitably spans multiple datacenter (DC) buildings. To support this scale, we designed a multi-building network that is capable of integrating hundreds of thousands of GPUs across nearby DC buildings into a single high-performance RoCE fabric. We first describe the network architecture within a DC building, followed by the architecture across multiple DC buildings.

\begin{figure}[t]
    \centering
    \includegraphics[width=0.4\textwidth]{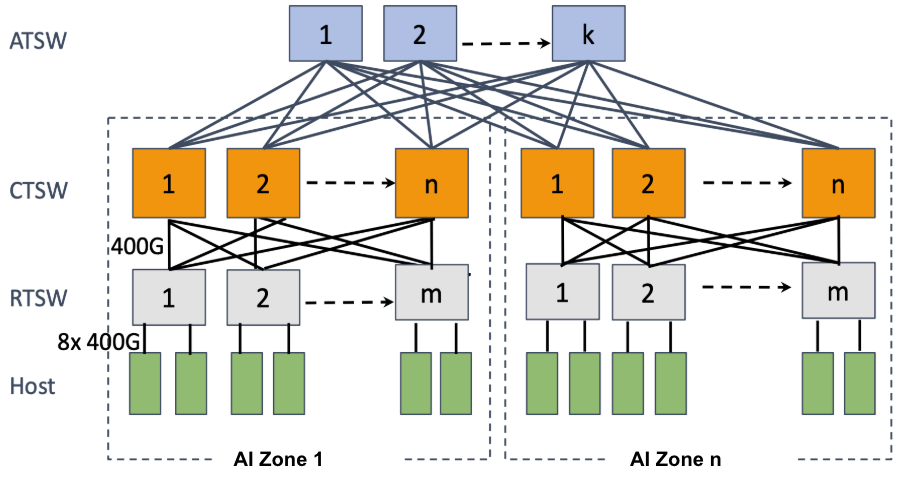} \\
    (a) Network architecture within a DC building. \vspace{1em}\\ 
    \includegraphics[width=0.4\textwidth]{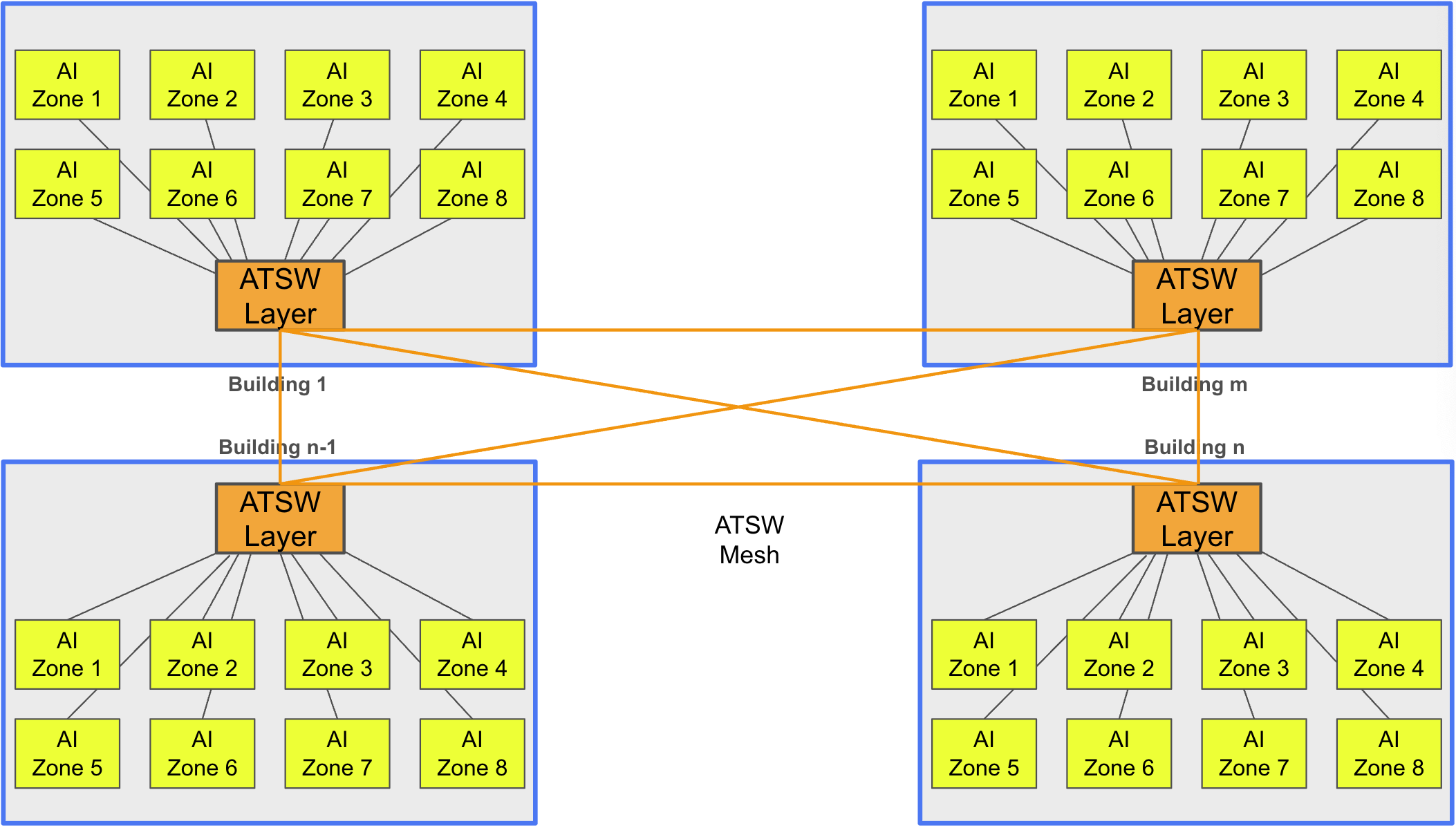} \\
    (b) Network architecture across DC buildings. \\
    \caption{Network architecture of the training cluster.}
    \vspace{-.1in}
    \label{fig:building_network_architecture}
\end{figure}

The network within a DC adopts a 3-layer Clos architecture~\cite{gangidiRDMA}, as shown in Figure~\ref{fig:building_network_architecture}(a). Each DC is partitioned into multiple AI Zones. The Rack Training Switch (RTSW) connects GPUs within a rack, while the Cluster Training Switches (CTSW) connect all racks within an AI Zone. Aggregator Training Switches (ATSW) connect CTSWs across the DC, extending the RoCE network beyond a single AI Zone. 
The cross-AI-Zone over-subscription ratio is 1:2.8.

To interconnect multiple DCs, we use a fully connected mesh between the ATSW layers of different DC buildings, as shown in Figure~\ref{fig:building_network_architecture}(b). Inter-DC traffic experiences the same over-subscription ratio as cross-AI-Zone traffic (1:2.8). This architecture is extensible and can scale to hundreds of thousands of GPUs within the same RoCE fabric, with DCs added incrementally over time.

In such a topology, GPU-to-GPU communication latency increases significantly as the number of network hops grows. Specifically, GPUs in the same rack have the lowest latency, while those in different racks within the same AI zone, in different AI zones, and in different datacenter (DC) buildings experience 7×, 15×, and 30× higher latency, respectively. 

Such a networking topology naturally motivates a replicated data-parallel design, which is adopted by \sys: It places all GPUs of a replica within one AI Zone or DC, so that the latency-sensitive
collectives within a replica do not need to operate over cross-DC links; it places different replicas in different DCs, since data-parallel collectives across replicas are more resilient to higher network latency.

\subsection{Hardware Reliability}

\begin{table}[t]
    \centering
    \begin{tabular}{lccc}
    \toprule
    \textbf{Component}  & \textbf{Count} & \textbf{\%} \\
    \midrule
        GPU HBM3 Memory  & 155 & 22.9\% \\
        PCIe Device  & 122 & 18.0\% \\
        NCCL Watchdog Timeouts  & 61 & 9.0\% \\
        Faulty GPU Compute  & 50 & 7.4\% \\
        Software Bug  & 48 & 7.1\% \\
        Host Maintenance  & 42 & 6.2\% \\
        Kernel Fault  & 39 & 5.8\% \\
        System Reboot  & 38 & 5.6\% \\
        Numerics/Silent Data Corruption  & 37 & 5.5\% \\
        Network Switch/Cable & 36 & 5.3\% \\
        SSD  & 30 & 4.4\% \\
        GPU SRAM Memory  & 8 & 1.2\% \\
        Unknown  & 7 & 1.0\% \\
        System Memory  & 2 & 0.3\% \\
        System Cooling  & 2 & 0.3\% \\
        GPU Thermal Interface + Sensor  & 1 & 0.2\% \\
    \bottomrule
    \end{tabular}
    \caption{Classification of unexpected interruptions during LLM pre-training on 32K GPUs.}
    \vspace{-.1in}
    \label{table:job_interruptions}
\end{table}

This section presents failure data from a recurring training job with about 32K GPUs.
It achieved between 95\%-97\% effective training time, with some training
time lost due to failures. On average, training experienced 2.3 interruptions per 1,000 servers
per day. 
Table~\ref{table:job_interruptions} summarizes
all training interruptions during our training on 32K GPUs. 
78\% of interruptions were due to hardware-related failures.
Among them, ``Faulty GPU Compute'' used to be the top cause, but is now reduced to the
fourth (7.4\%), 
thanks to our prior effort to address specific failure modes and the aggressive isolation of misbehaving GPUs.

As GPU compute issues declined, High Bandwidth Memory (HBM) issues rose to become the 
top issue in our current version. 
We see to improve the reliability of HBM subsystem in the future.

PCIe device issues were negligible in the previous version but became significant in the current version.
An increasing number of failures with SSD devices were due to a fatal NVMe firmware bug. 
This resulted in SSDs entering read-only mode with critical I/O errors. The other category 
of PCIe device issues involved GPUs, which reported uncorrectable errors, elevated correctable 
errors, and parity mismatch issues. In both cases, automation was able to detect and remove the faulty servers.

Although the Linux kernel is generally reliable, we observed occasional issues due to kernel faults, lockups, and spurious system hangs. These were typically transient and resolved by rebooting the affected servers. They may have originated from hardware or software bugs, or been triggered by specific workloads. Existing automation handled these cases effectively.

Although many reliability issues caused training jobs to restart, they were often well handled by our automation. However, training interruptions caused by numerical issues---such as Not-a-Number (NaN) errors or silent data corruption (SDC) detected through model evaluations---often required manual investigation and were difficult to debug~\cite{Wang2023SDC,RobustLLM}. The 37 interruptions in this category were traced to seven hosts, with root-cause investigations lasting from several hours to multiple days. In one SDC case, we traced the issue to 32 consecutive parameters (out of 2T total) in a single expert of a specific layer that abruptly spiked by over 1e7. We identified the faulty host by reversing the model-to-server mapping and using diagnostic tools. Additionally, we developed an internal tool to run deterministic training on servers and compare 
the results to detect those that occasionally experience SDC.

\vspace{-.1in}
\paragraph{Deterministic Training.}
To solve the SDC problem, we implemented a deterministic training flow to run on a single host to look beyond job termination with NaN for problematic hosts. Deterministic training ensures the following invariants:
1. The data for any training iteration is the same across runs.
2. All operators within the training flow give deterministic output for the same input. This is achieved by disabling flash attention optimizations; forcing cuDNN to use deterministic algorithms; and disabling cuDNN's auto-tuner.  

By doing so, we can generate equivalent checkpoints after a fixed number of iterations of the training loop across any number of runs of training. Any checkpoint maintains the model state for each rank. By comparing the checkpoint for the same ranks across multiple runs, we can find outliers when the checkpoint is not equivalent across a population.

\subsection{Fast Root Cause Analysis} 
\label{sec:rca}

To reduce training downtime, we must quickly identify the misbehaving server or GPU among tens of thousands of GPUs. Due to the synchronous nature of training, such issues often manifest as hangs or performance degradation in the collective communication library. We collect rich telemetry from the library and built a tool to accelerate root cause analysis. This tool periodically (every 5–20 seconds) fetches tracing data from all GPUs and performs deep analysis to pinpoint the source of the issues.

When a training job hangs without making progress, we first construct a wait-for graph of collectives. The leaves of this graph represent the collectives in which the entire job is blocked. For each of these collectives, we use telemetry to build a second wait-for graph, this time of GPUs within the collective. From the telemetry, we can determine whether GPUs failed to join the collective or joined but did not make progress. The leaves of this rank-level wait-for graph represent the specific ranks the job is ultimately waiting on.

If a training job continues to make progress but at significantly reduced performance, we aggregate the collective completion times for each rank. GPUs with substantially lower aggregate times are typically the ones causing delays for all others, as lower completion times indicate the GPU joins the collective last. Our approach leverages high-level communication-aware telemetry to identify faulty hosts—unlike prior work~\cite{deng2025minder}, which does not utilize such higher-level insights.

Our root cause analysis tool runs continuously, with each full analysis completing in under 5 seconds---even for jobs running on 32K or more GPUs. The telemetry data temporarily reside in host DRAM, not GPU HBM, and typically consume less than 1 MB per GPU in production. Data collection is asynchronous and incurs no impact on training performance.

To validate the accuracy of our tool, we injected approximately 1,500 failures across dozens of training jobs, each using hundreds of GPUs over multiple days. These failures caused the jobs to hang. Our tool correctly identified the offending server in 97.8\% of the cases. In the remaining 2.2\%, the faulty server was included in a small set of final candidates identified by our tool, which still significantly accelerates diagnosis. These rare inaccuracies typically occur when it is difficult to precisely attribute the issue to either the sender or the receiver side of a collective.

\section{Challenge: Scaling to 100K GPUs}
\label{sec:motivation}

As mentioned earlier, the paradigm of synchronous training and recovery
works fine with 32K GPUs, as we are losing about
3-5\% of GPU time due to the checkpoint-based recovery. However, when we try to scale to
100K GPUs, both the failure rate and the recovery cost increase,
significantly increasing the downtime.

\begin{figure}[t]
\centering
    \includegraphics[width=0.45\textwidth]{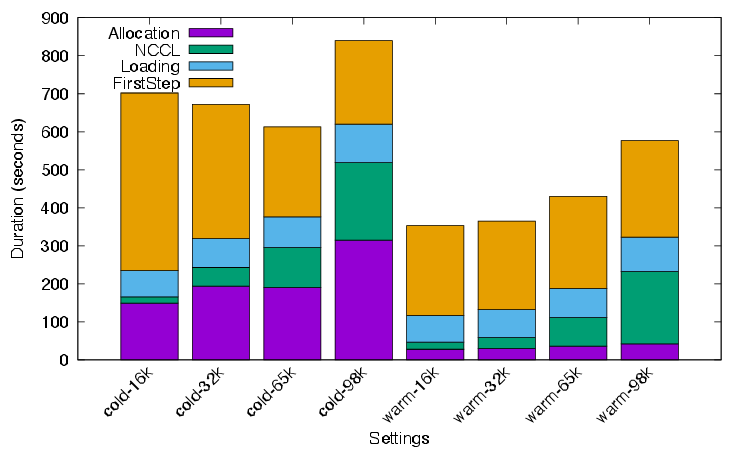}
    \caption{Breakdown of recovery time as the scale increases.}
    \label{fig:recover-time}
\end{figure}

Figure~\ref{fig:recover-time} provides a detailed breakdown of the time spent in different steps
of the recovery process.
Note that, since some of the steps are highly nondeterministic, ideally we should run
multiple experiments to filter noise, but since the large-scale experiments are
extremely resource intensive, we don't have the luxury to do so.

\vspace{-.1in}
\paragraph{Finding a replacement GPU (Allocation).} In the naive version (``cold-$N$'' in Figure~\ref{fig:recover-time}),
the system will stop the whole training job, return all GPUs to our private cloud, and ask
the cloud to restart the whole job. This step could take up to five minutes. As an optimization,
we can reserve a few GPUs as standbys and use them to replace failed GPUs when failure
occurs. This approach (``warm-$N$'' in Figure~\ref{fig:recover-time}),
which is also adopted in previous work~\cite{RobustLLM,comanici2025gemini25pushingfrontier}, can significantly reduce the allocation
overhead as shown in the figure.

\vspace{-.1in}
\paragraph{NCCL initialization.} Restarting training requires re-initialization
of NCCL. We find that the time to start all connections in NCCL grows with the
number of GPUs, from 17 seconds with 16K GPUs to about 200 seconds with 98K GPUs.
Note that this is already based on our own customized NCCL implementation, which tries
to reduce the start up time.

\vspace{-.1in}
\paragraph{Other loading overhead.} This includes the overhead of starting PyTorch, fetching and loading checkpoint, health check, etc. 
As shown in the figure, this overhead does not grow significantly with scale.

\vspace{-.1in}
\paragraph{``First-step'' overhead.} The first step of training typically takes more
time for extra tasks such as creating the checkpointer, initializing the dataloader 
(which involves caching the next N batches), Just-In-Time (JIT) compilation, etc.
As shown in Figure~\ref{fig:recover-time}, the first step can take several minutes, while normal steps
take about 20 seconds on average. With the ``cold'' approach, the time spent in the first
step is highly nondeterministic, probably because the ``cold'' approach needs to rebuild more entities.

\vspace{-.1in}
\paragraph{Summary and analysis.}
Our conclusion is that, despite our heavy effort to optimize these steps,
it is unlikely that we can keep the recovery time stable at a larger scale.
Note that ideally, we expect the recovery time to decrease with scale to compensate for the higher
failure rate we will face. Concretely, our production data show that, with 100K GPUs,
there is a failure every 18 minutes. If the system
needs to stall for 10 minutes for recovery every 18 minutes, we can only achieve an effective
training time of about 44\%.

These problems naturally motivate us to explore the idea of asynchronous recovery. Asynchronous
recovery could hide the latency of allocation, NCCL initialization, and other loading
steps. It may not be able to fully hide the extra latency of the first step, i.e., the recovering
replica may be slower in the first step after it recovers, slowing down the entire system,
but we may be able to shuffle some of such first step overhead into the recovery process.

\section{Design of \sys}\label{sec:overview}

\begin{figure}
\centering
    \includegraphics[width=0.5\textwidth]{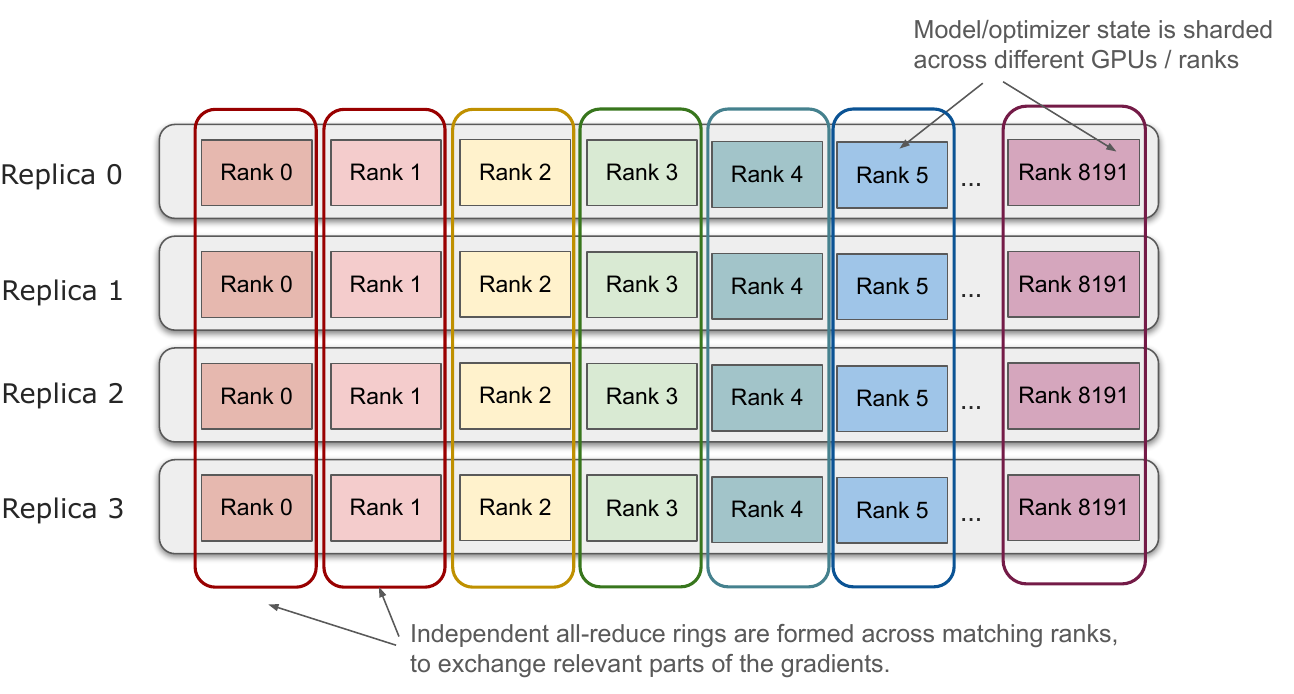}
    \caption{Architecture of \sys.}
    \label{fig:arch}
\end{figure}

To address the long and frequent recovery problem discussed above,
\sys adopts the HSDP paradigm. As shown in Figure~\ref{fig:arch}, it creates multiple
replicas, each processing a subset of training data.
Each GPU in a replica is given a unique rank number, depending on its position in the replica.
After training a fixed amount of data called a batch,
GPUs of the same rank from different replicas exchange
gradients. As discussed in Section~\ref{sec:networking}, GPUs of the same replica will be allocated from the same DC,
and gradient exchange will be executed across DCs.

This paradigm brings two benefits for the purpose of fault tolerance. First, if a GPU fails, \sys only needs to rebuild
the replica that includes the failed GPU, which reduces the
scale of recovery and thus reduces the recovery time. 
Second, other replicas can still continue training while the failed replica is recovering.
Of course, losing a replica will reduce overall training throughput, but
this is much better than stalling the entire system during recovery.

This section first presents an overview of \sys and
then discusses its key techniques in detail.

\subsection{Overview}
\label{sec:workflow}

\begin{figure}
\centering
    \includegraphics[width=0.5\textwidth]{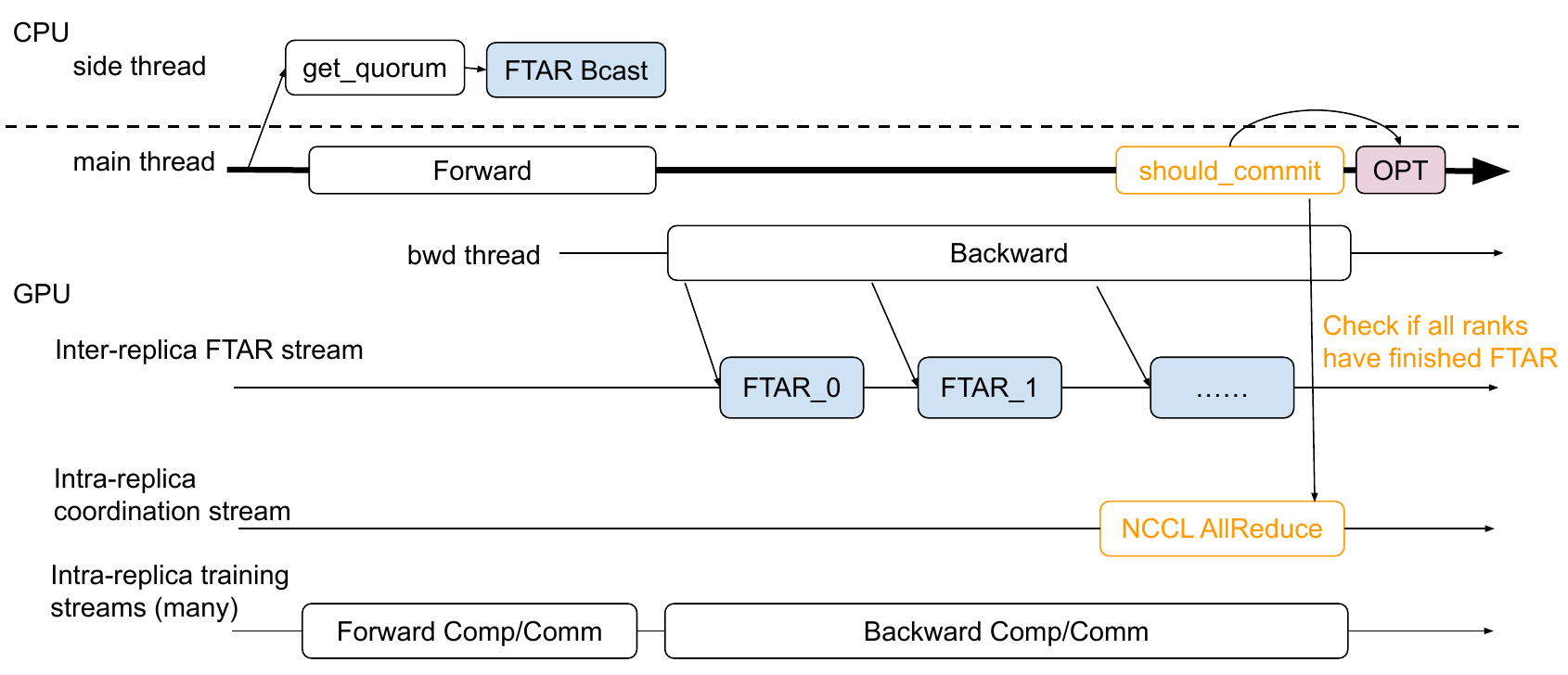}
    \caption{Design of Rank 0 (leader) in \sys.}
    \label{fig:PAFT-details}
\end{figure}

Following the idea of Hybrid-Shared Data Parallelism (HSDP), \sys works in multiple ``steps''. In each step, each replica performs
training on its own batch of training data, involving forward and backward computation and communication.
Then GPUs of the same rank exchange gradients to compute the sum of their gradients.
Finally, the optimizer at each rank applies
the gradients to update its model weights.
A GPU may perform a checkpoint at the end of a step to write its 
own state to persistent storage (see Section~\ref{sec:checkpointing}).

Figure~\ref{fig:PAFT-details} shows the detailed design of Rank 0, which is the leader
of a replica. Like other ranks, it includes multiple intra-replica streams to perform forward
and backward computation and communication; it includes one inter-replica
stream to perform gradient exchange (i.e., FTAR-related, which will be discussed in Section~\ref{sec:ftar});
its main thread executes the optimizer step (OPT in the figure) to apply gradients to update model weights.
Note that \sys overlaps FTAR with backward computation to improve training speed.
As the leader of a replica, Rank 0 needs to execute extra control logic: 1) A CPU
thread coordinates with Rank 0s in other replicas through a consensus service
(similar to Chubby and Delos~\cite{Burrow2006Chubby,Balakrishnan2020Delos}) to determine healthy replicas.
2) Rank 0's main GPU thread will ask other ranks in the same
replica to determine whether they have successfully finished gradient exchange. If so,
they can proceed to the optimizer step to update model weights.

\vspace{-.1in}
\paragraph{Failure detection.}
The implementation of \sys described herein relies on timeout to detect failed GPUs for simplicity.
One step in training takes about 20 seconds on average, so empirically, we set our timeout
interval to be 60 seconds. In the future, we plan to integrate with the fast root cause
analysis component (Section~\ref{sec:rca}), which hopefully will further reduce stall time.

\vspace{-.1in}
\paragraph{Ensuring consistency after failures.}
For training quality, ideally all GPUs should be in a consistent state, i.e., their trained models should
correspond to the result of training the same input data. Synchronous training systems achieve this
by restarting all GPUs with the lastest checkpoint when
a failure occurs.

However, in \sys, when a GPU fails,
other replicas may not be in a consistent state. For example, if Rank 0 of Replica 0
in Figure~\ref{fig:arch} fails during a step, it may happen that, before failure, Rank 0 of Replica 0
has successfully exchanged gradients with Rank 0 of Replica 1, but not with Rank 0s of Replica 2 and Replica 3.
This can create inconsistency both within a replica and between replicas.

A naive solution to this problem is to ask every replica to retrain the failed step,
but this approach may cause a waste of resources when some replicas have successfully completed
their gradient exchanges.
For a more efficient solution, our idea is that while consistency within a replica is
necessary, consistency across replicas is unnecessary due to \sys's support for asynchronous recovery:
If one replica completes a step and another one does not, \sys lets the complete replica
continue training; \sys lets the incomplete replica re-join later, like a recovering
replica.

Concretely, after a failure is detected, GPUs within a replica will run a protocol like two-phase commit (2PC): 
Rank 0 of the replica will ask every other rank in the replica whether it has finished gradient exchange (Figure~\ref{fig:PAFT-details});
If everyone replies yes, Rank 0 will ask everyone to apply the gradients to update its model weights (i.e., optimizer step)
and proceed to the next step.
Otherwise, Rank 0 will ask all the ranks in the replica to retry the step. Note that in this case, all the ranks
have not updated their model weights yet, so they can simply discard their gradients instead of fetching
model weights from others.

Since each replica makes such a decision independently, they may end up in different situations. In the
previous example, Replica 1 may decide to proceed to step 100, and Replicas 2 and 3 may decide to retrain
step 99. In this case, Replicas 2 and 3 will trigger the recovery protocol.

\vspace{-.1in}
\paragraph{Asynchronous recovery.}
\sys recovers a replica while the other replicas continue training.

As discussed above, recovery can occur in two cases.
First, for a replica with failed GPUs, \sys will find replacement
GPUs, re-build the replica, reboot all GPUs, and run a protocol to
let the recovering replica join others. We have developed a
non-blocking catch up protocol to minimize the stall during
the join process (Section~\ref{sec:catchup}).
Second, due to possible inconsistency after a failure discussed above, some replicas
may be left behind. These replicas can execute the same
non-blocking catch up protocol to join others.

\subsection{Efficient Fault Tolerant All Reduce}
\label{sec:ftar}

\sys uses NCCL for intra-replica communications due to its superior performance, but introduces
its own fault tolerant all reduce (FTAR) protocol to exchange gradients
across replicas since NCCL cannot achieve all the goals of FTAR:

\begin{itemize}

\item Network-aware architecture. FTAR needs to perform across 
DCs, which means that it cannot involve any NVLink-domain communication. 
Since bandwidth across DC buildings is oversubscribed (Section~\ref{sec:networking}), FTAR must
ensure high speed with congestion control in mind.

\item Reconstructable communication group. When a fault occurs, we require each FTAR group to be reconstructed on the healthy ranks. That is, the healthy rank can clean up existing stale transport resources and re-establish the transport connection with the new group of peers.

\item Easy fault management.  When a replica fails, the GPU ranks in the healthy replicas should stay alive
and keep the training state in memory. To achieve it, FTAR needs to be able to detect
whether an error is recoverable or fatal. If it is recoverable
(e.g., network remote error due to a dead peer), then FTAR can report the error to
\sys and let \sys remove the dead pear. If it is fatal (e.g., bad NIC),
then FTAR should instruct \sys to kill the rank.

\item Minimal resource contention to concurrent computation kernel. \sys overlaps FTAR with the backward computation, so we need to ensure minimal GPU resource contention to the concurrent computation kernels.
\end{itemize}

These goals are not difficult to achieve in a traditional CPU-based protocol.
However, we find that NCCL is not sufficient to achieve these goals. One problem with NCCL
is that its communication is driven by the GPU, but as a device primarily designed
to accelerate computation, GPU still lacks CPU's capability to implement complex
logic, such as treating different kinds of errors separately, reconstructing the FTAR
group, congestion control, etc.  As a result, we adopt a hybrid design
that let CPU drive the whole process.

\begin{figure}[t]
\centering
    \includegraphics[width=0.5\textwidth]{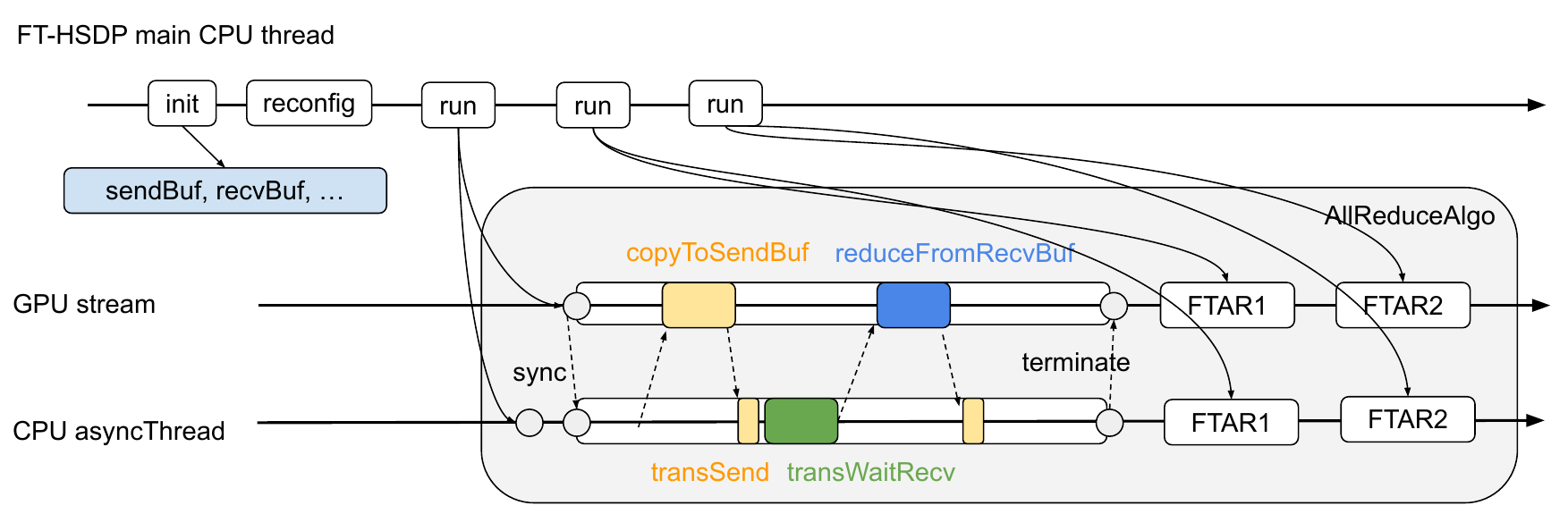}
    \caption{Details of FTAR protocol.}
    \label{fig:ftar-details}
\end{figure}

\vspace{-.1in}
\paragraph{CPU-GPU interaction.}
Figure~\ref{fig:ftar-details} presents how CPU and GPU interact in FTAR.
The \sys main thread, which runs on CPU, will initialize the RDMA connections by creating
send buffers, receive buffers, etc. It will also run the
reconfig function, which determines which ranks will join the FTAR group. Then
it runs FTAR by starting a GPU stream and a CPU asyncThread: The GPU stream copies
data to the send buffer and notifies the CPU asyncThread; the CPU asyncThread asks the
GPU to send the data in the send buffer to the remote GPU rank through RDMA; the CPU
asyncThread also waits for data from other ranks;
then the asyncThread notifies the GPU stream that data have arrived; the GPU stream
runs the reduce function on the receive buffer; and so on. \sys may run FTAR for
multiple iterations, each exchanging a portion of the gradients for the purpose of
congestion control.

This hybrid design utilizes the CPU for the control plane: The CPU maintains
connections so it can destroy them or create new ones freely; the CPU can determine the peers
by communicating with other CPUs through a consensus service; the CPU can limit the amount of data in flight for congestion
control; and the CPU can determine how to handle an error depending on the error type.
Such complex control logic is hard to implement fully in GPU. On the other hand, this design utilizes
the GPU for the data plane: All data transfer is executed by the GPU to maximize performance.

\vspace{-.1in}
\paragraph{Ring algorithm.} In the use case of FTAR, we expect that participanting GPUs
are most likely distributed across AI zones and across buildings. The zone and building switches
exhibit high oversubscription ratios and limited bisection bandwidth compared to those within a zone (Section~\ref{sec:networking}).
Therefore, it is important to avoid potential network congestion across zones or buildings. 
To achieve this, the number of concurrent data packets transmitted 
within each FTAR group should be restricted. We estimate that FTAR 
should be optimized for message sizes ranging from 200MB to 500MB, accommodating up to 16
ranks. Such a scenario often prefers bandwidth-optimal algorithms. 

Based on these conditions, we choose the ring algorithm that limits each GPU to only sending and
receiving with two neighbors in a Ring, thus minimizing the concurrent network traffic within 
each FTAR group. We adopt the ring algorithm~\cite{Patarasuk2009Bandwidth,Thakur2005MPIOptimization}
consisting of a ReduceScatter phase and an AllGather phase, 
each containing N-1 steps (N is the number of GPUs in the Ring = the number of replicas):
In the ReduceScatter phase, the GPU of replica i starts sending the i-th portion to its right neighbor. Each GPU
performs an in-GPU reduction upon data arrival, and then forwards the reduced data to its right neighbor.
After the last step of the ReduceScatter phase, each GPU should have the final result of a portion of the data. 
In the AllGather phase, each GPU forwards its final result portion to its right neighbor. Upon data 
arrival, the GPU then forwards the received result to the right neighbor. Similarly, all GPUs 
will eventually receive all the portions in the last step.

Following the ring workflow, we then design the pipeline protocol with fixed-size chunks for varying FTAR message sizes. Specifically, we pre-allocate an internal sendBuf and a recvBuf on each GPU when initializing FTAR resource, each with a fixed-size chunk size (S) $\times$ number of chunks (C). The allocated buffers are pre-registered to the network for RDMA transmission.  For a given FTAR group with N ranks, the data is split into multiple partitions, each with at most S $\times$ C $\times$ N bytes. Within each partition, FTAR finishes one round of Ring algorithm with 2N-2 steps, and pipelines with C chunks. Once the first partition is complete, FTAR then performs the next partition, and so on.

The fixed-size chunk based pipeline brings two benefits: First, it controls the amount of concurrent data packets between every two peers to be at most S $\times$ C bytes. Second, it provides flexible performance tuning for the copy/reduce kernel operation and the network transfer operation, thus allowing us to tune each operation to achieve optimal performance. Specifically, for a given chunk size, developers can separately tune the number of thread blocks to adjust the throughput of the in-GPU copy/reduce, and tune the number of Queue Pairs and other transport-specific hyperparameters for the throughput of network transfer.

\vspace{-.1in}
\paragraph{Optimizations.} We optimize the reduce and copy kernel operations in several ways. 

\begin{itemize}[leftmargin=*]

\item In the ReduceScatter phase, steps 0 to N-2 require the reduction result to be stored only in sendBuf; in step N-1, the result is stored in both the local data buffer (as the final result returned to user) and the sendBuf. Similarly, for steps 0 to N-2 in the AllGather phase, the received data must be stored to both local data buffer and the sendBuf; the last step of the AllGather phase copies data only to the local data buffer.

\item The last step of ReduceScatter needs to store data to both the sendBuf and the local data buffer.
We combined these two steps, which reduces one round of CPU-kernel synchronization compared to handling them as separate kernel
operations. Moreover, the combined approach avoids the HBM load in the later step, since the results are still in the register after each reduction instruction. 

\item Our algorithm launches a CUDA kernel for each AllReduce operation, which performs busy-polling on a flag in host-pinned memory. This busy-polling can waste GPU resources while the kernel waits for a CPU thread to signal that communication is complete, allowing the kernel to proceed with copy and reduction operations. This approach is common in communication libraries like NCCL, as it reduces the overhead associated with kernel launches. To minimize the wastage of GPU resources, the kernel is typically launched on a small number of Streaming Multiprocessors (SMs). For example, on the H100 GPU, which has 132 SMs, NCCL AllReduce utilizes only 4 SMs. As a result of this algorithm, it is necessary to develop kernels that can achieve high memory bus utilization and performance while operating with low occupancy. To address this, we leverage instruction-level parallelism by issuing multiple memory load/store and reduction instructions per thread. This requires a large number of registers to be available per thread. Consequently, we implemented several optimizations to reduce the number of registers used: We optimized unnecessary algorithm context information, performed as many computations as possible at compile time using constexpr expressions, and reduced the block size. Collectively, these optimizations allowed us to reduce the kernel's copy and reduction operations, effectively hiding them behind communication for the target message size.
\end{itemize}
Putting together all the optimizations, FTAR uses two thread blocks (i.e., SMs), 512 threads per block,
and 8MB chunk size to fully hide the reduction within concurrent network transfer. It uses even less GPU compute resources (SMs) than that needed by native NCCL AllReduce (FTAR uses 2 SMs whereas NCCL uses 4 SMs). We note that there is no gain to further reduce the reduction/copy overhead if we need to pay more SMs, since AllReduce is already network-bound. In contrast, using more SMs in the AllReduce may significantly degrade the performance of the other compute kernels. 

\vspace{-.1in}
\paragraph{Timeout and error handling.}
When the GPU stream needs to wait for data or signals from a peer, we use a while loop to check
a flag with a timeout to implement the wait logic, ensuring that the GPU will not hang.

We categorize errors into two types: recoverable errors and non-recoverable errors, and react to them differently.
FTAR retries a recoverable error, which includes timeouts and network errors (i.e., ncclRemoteError). Concretely, when FTAR catches a recoverable error, it will return a retryable work back to \sys so that the upper layer control plane can retry with it. Since the following GPU operations might have been queued already, FTAR will also terminate/cancel them and mark them as skipped.
FTAR quits immediately when getting a non-recoverable error, which usually indicates a bug or hardware failure, such as ncclSystemError, ncclInvalidArgument, etc. Retrying these errors would only waste time, so we let FTAR crash according to the error type.

\vspace{-.1in}
\paragraph{Summary.} As shown above, FTAR relies on the CPU to perform complex control logic,
relies on the GPU for the actual data copy, and incorporates various optimizations to maximize performance.
As shown in Section~\ref{sec:eval-ftar}, FTAR can achieve a throughput comparable to native NCCL.

\subsection{Non-blocking Catch Up}
\label{sec:catchup}

\begin{figure}[t]
\centering
    \includegraphics[width=0.5\textwidth]{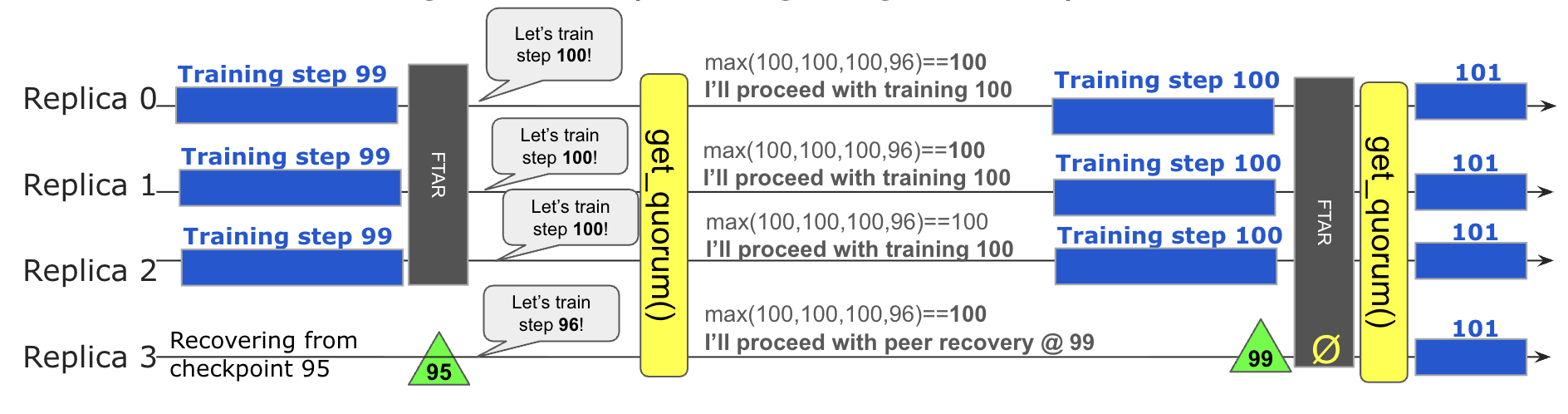}
    \caption{Non-blocking catch up.}
    \label{fig:catch-up}
\end{figure}

A replica may fall behind other replicas for multiple reasons. For example,
a recovering replica has random model and optimizer states after initialization and thus has to retrieve
these states from other replicas. As discussed in Section~\ref{sec:workflow}, after a replica
fails, the remaining replicas may end up in different states.

In a traditional replicated system, the fall-behind replica needs to fetch
the latest checkpoint from other replicas and then executes tasks since that checkpoint.
However, this process may face the challenge of the ``catch up'' problem: The other
replicas are executing new tasks (training new data in the case of \sys) while the recovering replica is fetching the checkpoint,
so the recovering replica is still falling behind after it loads the checkpoint. To 
ensure that it can catch up with others, we typically need to stall or slow down the other
replicas, which is not desired .

Instead, we propose a protocol that leverages the specific properties of training to achieve
non-blocking catch-up. Concretely, at the beginning of a step, each replica reports its next step number through
a consensus service. The replicas with the highest step number $n$ are considered healthy and proceed
to train step $n$. The other replicas are left behind and thus fetch the checkpoint corresponding to the end of step $n-1$
from the healthy replicas. At the end of the step, healthy replicas send their gradients as normal, and the left-behind replicas
send a zero gradient. The gradient exchange protocol (i.e., FTAR) will automatically bring all replicas to the same state.

Figure~\ref{fig:catch-up} shows an example. At the beginning, Replica 3 is falling behind
and other replicas are healthy. When they exchange information, Replicas 0-2 report that
they are going to train step 100 and Replica 3 is at step 96. As a result, Replicas 0-2 know
that they are healthy and should proceed with training 100; Replica 3 finds that it
is falling behind and thus should fetch the checkpoint corresponding to step 99 from other replicas.
At the end of the step, Replicas 0-2 send their gradients and Replica 3 sends a zero gradient.
FTAR will bring all replicas to the same state and then they will train step 101 together.

Compared to recovery in a general-purpose replicated system, two properties of
the training system make such non-blocking catch up possible. First, as long
as all replicas have the same checkpoint, a replica which does not do any training work
can send a zero gradient and then reach the same state with other replicas which did the training work. In
a traditional replicated system, however, the fall-behind replica needs to actually execute
the tasks. 
Second, due to
our efficient checkpoint recovery mechanism (Section~\ref{sec:checkpointing}), fetching a checkpoint usually
takes shorter than a step, so a healthy replica does not need to wait for
a fall-behind replica. Of course, if fetching checkpoint is slower than training a step
for some reason, then the training replica needs to wait, causing blocking, but this
is very rare in our experience.

\subsection{Efficient Checkpoint and Recovery}
\label{sec:checkpointing}

\begin{figure}
\centering
    \includegraphics[width=0.5\textwidth]{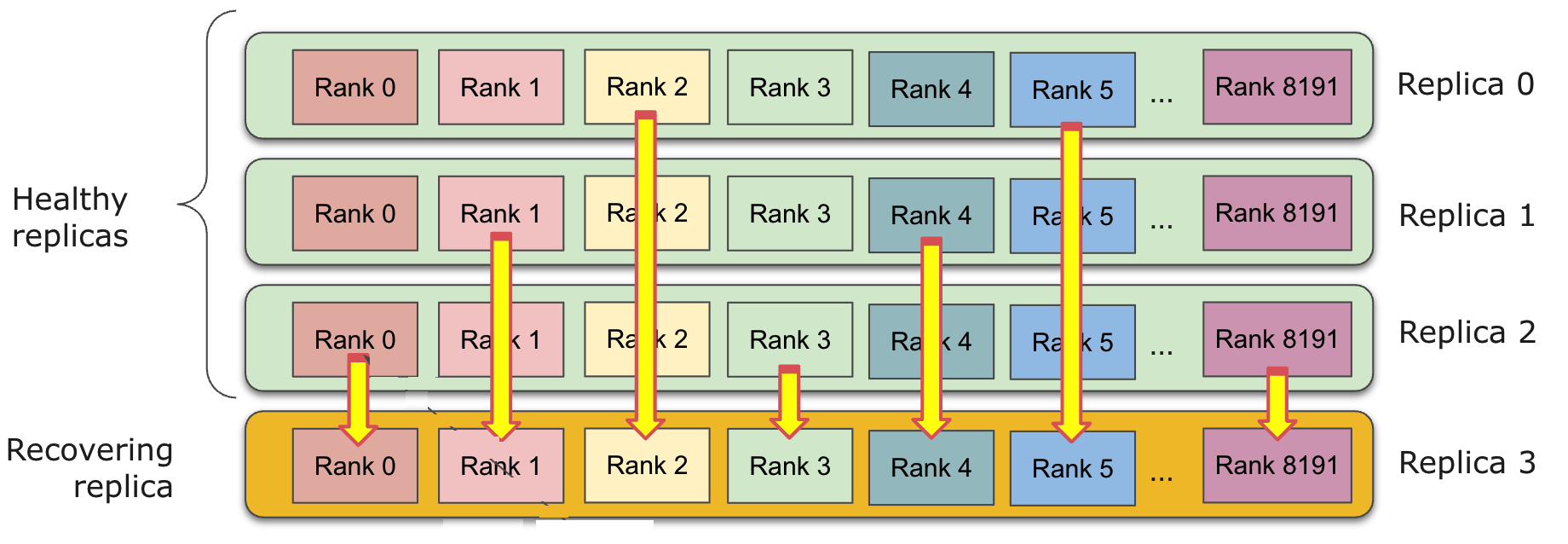}
    \caption{Fetching checkpoint.}
    \label{fig:checkpoint}
\end{figure}

In the common case, a GPU will write its state to a remote persistent
storage every 100 steps.
Concretely, a GPU's state includes model state, optimizer state, and dataloader
state. Since model state and optimizer state are replicated across
replicas, only one replica needs to write those states into the persistent
storage. The dataloader state, which marks the position of the data
to be trained, is not replicated, but it is very small, so \sys writes
the dataloader state of all GPUs into a single loader\_state file every step.

The persistent checkpoint is used for full system reboot.
However, during the catch up protocol, the recovering replica
will fetch state from healthy replicas directly~\cite{Wang2023AmazonGemini}, which allows
the recovering replica to fetch state corresponding to the latest step.
Since states
are replicated, the recovering replica only needs to fetch from one,
and \sys organizes such fetching in a load-balanced manner (Figure~\ref{fig:checkpoint}),
similar as it does for persistent checkpoint. In this process,
the healthy GPU will write its state to CPU memory, and then the recovering
GPU will fetch the state through an HTTP protocol, which has
no contention with the forward and backward communication through
the GPU's high-speed network. The recovering replica will recover
its dataloader state from the loader\_state file mentioned above.

This process assumes that a GPU will not change its state while another
GPU is fetching its state, which is true before the optimizer step. Note that
this may not be true for a general-purpose distributed system, so if the system
wants to fetch state while executing new tasks, it often involves complex
techniques like Copy-On-Write (CoW).

\section{Implementation}
\label{sec:impl}

\vspace{-.05in}
\paragraph{Train each batch exactly once.}
As a criteria set up by the ML experts, \sys should not
skip training data or train the same batch more than once.
Following this criteria, a failed replica will resume
from exactly where it fails with the help of the loader\_state file. 

This strategy may cause a replica to become a straggler at the end. However,
our current experience is that this is not a severe problem, as severe straggler
only happens if one replica is constantly having more failures than others, which
is rare. More sophisticated strategies, such as reshuffling batches when a failure
occurs, are possible.

\vspace{-.1in}
\paragraph{Reducing Memory Pressure.}
For the purpose of fault tolerance, we want to run a large number of small replicas,
so that we don't lose much training capacity due to one failure. However,
this strategy creates significant memory pressure as each replica has to contain
all states.
We mitigate this problem in two ways. First, we test different number of replicas to
find a good balance. So far, we are settled with 10-20 replicas.

Second, we implement a mechanism to offload optimizer state from GPU memory to
CPU memory. And the GPU fetches
such state from CPU memory on demand. Our evaluation shows that a GPU needs about 60ms to fetch
1GB of optimizer state from the CPU memory.

\vspace{-.1in}
\paragraph{Reducing the first-step effect.}
As discussed in Section~\ref{sec:motivation},
the first step after a recovery on a replica involves some time-consuming initialization functions, such as re-creating the checkpointer and initializing the dataloader. As a result, in the first step after recovery, the recovering replica may be slower than healthy replicas,
causing stalling.
We mitigate these issues by asking a recovering replica to execute these initialization functions prior to checking in with \sys, when possible.
One notable exception is loading the first batch, which must be done between the check-in and the first FTAR.
As a result, our implementation reduces but does not fully eliminate the first-step effect, as shown in Section~\ref{sec:eval}.

\vspace{-.1in}
\paragraph{Large-scale emulation with CPUs.}
Software testing is a critical component of any software project’s development process, but
for \sys, running tests on 100K GPUs regularly is not practical. Running tests with fewer
GPUs is helpful but may miss problems that occur only on a large scale.

To mitigate this problem, we have built a tool to emulate a large-scale training on CPUs,
since CPU resources are much more abundant. Concretely, we have developed a shadow
module loading mechanism, which transparently replaces default GPU targeted modules with CPU based mock modules.
For these mock modules, the exact implementation choices are case-by-case. In many cases, for computation,
we can skip the real computation and only pass along the tensor shape information to continue the high level training loop’s logical steps. 
For network communication, we use custom CPU based basic communication libraries to substitute the NCCL code path.

Of course, this tool does not help to find problems in GPU-related code, but since many components in \sys,
such as dataloading pipe and preprocessing functions, only run on CPUs, this tool helps to find
problems in these components. In addition, it also helps to test external services \sys depends on,
such as job scheduling, storage, telemetry publication, logging, etc.

This tool has helped us detect and fix many bugs in various components, 
all conducted well before the real GPU platforms became available,
To give an example, when testing an early version of get\_quorum, which is the consensus protocol to determine which
replicas are healthy (Figure~\ref{fig:catch-up}),
this tool reported that its latency is 9 seconds at 100K scale,
which would be a significant overhead. We iterated the fixes with this tool experiments and 
successfully lowered the latency to 700ms. And when the final GPU cluster became ready for the test,
the GPU run exhibited the same latency.

\section{Evaluation}
\label{sec:eval}

Our experiments try to answer the following questions: 

\begin{itemize}

\item How much stall
time does \sys introduce for failure and recovery? (Section~\ref{sec:eval-stall})

\item Does failure and recovery affect the accuracy or convergence of training? (Section~\ref{sec:eva-accuracy})

\item What is the performance of FTAR? (Section~\ref{sec:ftar})

\end{itemize}

Due to the high resource requirement of running experiments with a large number of GPUs,
we run one experiment with 98K GPUs to answer the first question.
We run smaller experiments to answer the second and third questions.
We present the detailed experiment setting in each corresponding section.

\subsection{Stall Time and Training Efficiency}
\label{sec:eval-stall}

We run this experiment with 98K H100 GPUs, which are partitioned into 12 replicas and spread across 4 datacenter buildings with the network topology presented in Section~\ref{sec:networking}. Each replica trains a dense transformer model with several trillion parameters on 8192 GPUs. We train each replica with fully sharded data parallelism, asynchronous tensor parallelism, context parallelism, and a custom pipeline parallel schedule on a combination of text and image tokens.

We start the experiment with 12 healthy replicas. They can reach a throughput of about 450 TFlops/GPU/s. This is the same as the steady-state throughout without \sys, indicating \sys has no overheads in the failure-free case.

When we kill a replica, it takes \sys about three minutes to detect and handle the failure and re-train the failed step.
This is more than our expectation, which includes the 60-second timeout to detect the failure, a few seconds to
rebuild FTAR, and about 20 seconds to re-train a step. Our analysis reveals two reasons. First, we have incorporated
a power smoother technique to prevent abrupt changes in power consumption. In our case, a failure stalling all 98K GPUs
will trigger the power smoother.
Our analysis shows that the power smoother implementation incurs a 45-second stall. We have since come up with a solution to prevent this stall. Second, we find a bug that causes 45 seconds of additional stall to perform unnecessary
FTAR reconfiguration. We have fixed the bug afterward. With these two fixes, the stall time should be around 1.5 minutes,
which is close to our expectation. We plan to rerun this test to confirm when 100K GPUs are available for training again.

Then \sys continues the training with 11 replicas at a throughput of about 450 TFlops/GPU/s, which is as expected.

When the failed replica rejoins \sys at step $n$, it takes \sys about two minutes to complete step $n+1$. 
Considering a step typically takes 20 seconds, this means that rejoining introduces about 100 seconds of stalling. There are two reasons
for this stall. The first is due to the ``first step'' effect as discussed in Section~\ref{sec:motivation}. Despite our effort to 
shuffle some of the extra tasks to execute before the first step,
we do not completely eliminate that  (Section~\ref{sec:impl}). Note that even if all 100 seconds are due to the first-step effect, it is shorter
than the about 200-second first-step time as shown in Figure~\ref{fig:recover-time}, which indicates that our optimization
is effective.
The second reason is that, if a replica joins in the middle or close to the end
of step $n$, its recovery may extend beyond the end of step $n$. In this case, \sys has two options: First, it could ask the recovering replica
to join at the end of step $n+1$, which causes no stall but could not utilize the replica in step $n+1$. Second, \sys could
ask other replicas to wait for the recovering replica at the end of step $n$, which causes stall, but could utilize the replica in step $n+1$.
To balance such a trade-off, we set a time limit for waiting, which may incur a stall. Both reasons are as expected.

Afterwards, \sys continues the training with 12 healthy replicas at a throughput of about 450 TF/GPU/s, which means it fully recovers.

In summary, \sys has a total stall of about three minutes for failing and recovering one replica.
We analyze the improvement as follows assuming one failure every 18 minutes: With the synchronous-recovery approach, the system will
stall completely for about 10 minutes every 18 minutes, which means the system can achieve an effective training time of only $\frac{18-10}{18}=44\%$. With
\sys, assuming that it uses 12 replicas, fully repairing a replica still takes 10 minutes, but it only
stalls completely for 3 minutes (i.e., running with 11 replicas for 7 minutes), then the effective training time of \sys is $\frac{8 + 7 \times \frac{11}{12}}{18}=80\%$.
Although not ideal, this is much better than the synchronous-recovery approach.

As discussed in Section~\ref{sec:workflow}, it is possible to further reduce this 3-minute stall by replacing the
60-second timeout with a faster failure detector.

\subsection{Model Accuracy} 
\label{sec:eva-accuracy}

For asynchronous recovery, an important question is whether frequently removing and adding
replicas will affect the accuracy of the result model.

Due to the resource burden of running experiments at full scale, we can only run the full-scale
experiments once with a short duration. During this short duration, we observe that the training progress of \sys
matches our expectation.
For a more comprehensive study, we run experiments on 256 H100 GPUs, training a number of MoE models with three billion active parameters and 16 experts. We use a training set with 500 billion diverse text tokens in all experiments. 

Of course, it is questionable whether we can extrapolate the conclusions from small scale experiments to the full scale.
However, considering that almost no labs or companies can afford to run
multiple experiments at full scale for comparison, it is a standard practice to perform hyperparameter tuning
and comparison at small scales and directly apply the best working recipe at full scale. This works well for us in the past
and we consider it the best we can do.

We test different settings. We represent each setting as $freq\_fp\_len\_con\_lr$. $freq$ is the number of failures per n steps (n varies across experiments); $fp$ is the floating point precision, either fp8 or fp16; $len$ is the length (i.e., number of steps) of a failure; $con$ is the number of concurrent replicas killed in each failure; $lr$ is the learning rate intervention strategy, which is discussed later. 
For highest $len$ value 4k , we set $n$ to be 11k. For other settings, we set $n$ to be 5k.
For example, $d2x\_fp8\_for4k\_1reps\_lr\_none$ means that the experiment introduces two failures per 11k steps, uses fp8 precision, lets each failure last 4k steps, kills one replica in one failure, and does not use any learning rate intervention strategy. 

Learning rate is an important parameter during training. Typical training systems, including our baseline system,
sets the learning rate to a high value at the beginning to accelerate convergence and decreases its value gradually based on step number
to avoid oscillations. In \sys, since different steps may train different amount of data due to failures, it's probably
better to take such failures into consideration when adjusting the learning rate.
Concretely, we test three strategies: $none$ uses the baseline's step-based learning rate adjustment. $linear$ intervention scales down the baseline's learning rate by $\frac{num\_healthy\_replicas}{num\_total\_replicas}$, since the amount of data to be trained in a step is determined by the number of healthy replicas.
$sqrt$ intervention scales down the learning rate by $\sqrt{\frac{num\_healthy\_replicas}{num\_total\_replicas}}$, where we ensure the learning rate is always proportional to the standard deviation of the gradient noise.
$sqrt$ performs better than $linear$ in our experiments in general, so we do not show $linear$ results.

We measure the progress of the training with both the classic loss function (negative log likelihood of predicting the next token) and four external evaluation sets containing coding, reasoning, math, and general texts not included in the training data.
Due to space limit, we present a subset of the results and the conclusion holds in other settings.

\begin{figure*}[t]

\centering
     \begin{subfigure}[b]{0.45\textwidth}
     \includegraphics[width=\textwidth]{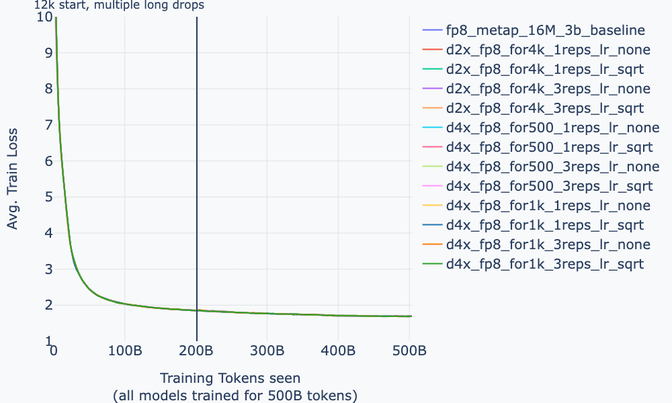}
     \caption{Training progress measured by loss function.}
     \end{subfigure}
     ~
     \begin{subfigure}[b]{0.45\textwidth}
     \includegraphics[width=\textwidth]{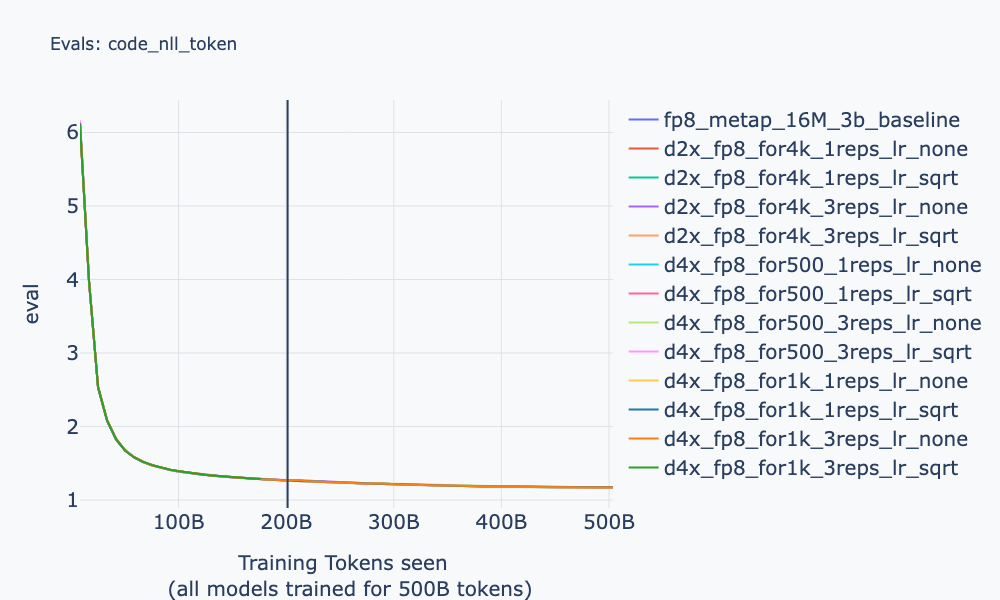}
     \caption{Training progress measured by the external coding set.}
     \end{subfigure}
     \caption{Overall training progress (we start killing replicas at 200B tokens).}
     \label{fig:accuracy1}
\end{figure*}

Figure~\ref{fig:accuracy1} shows the full training progress, measured by both the training loss and the coding evaluation
set (the other three evaluation sets show the same trend). In overall, we do not observe a distinguishable difference between
the baseline and the different settings of \sys, even for the most aggressive failure scenario (i.e., $d2x\_fp8\_for4k\_3reps\_X$).
Note that these figures use the number of tokens seen as the x-axis. If we use the running time as the x-axis, \sys
will take longer to reach the same progress as the baseline since \sys needs to handle failure recovery.

\begin{figure*}[t]
\centering
     \begin{subfigure}[b]{0.45\textwidth}
     \includegraphics[width=\textwidth]{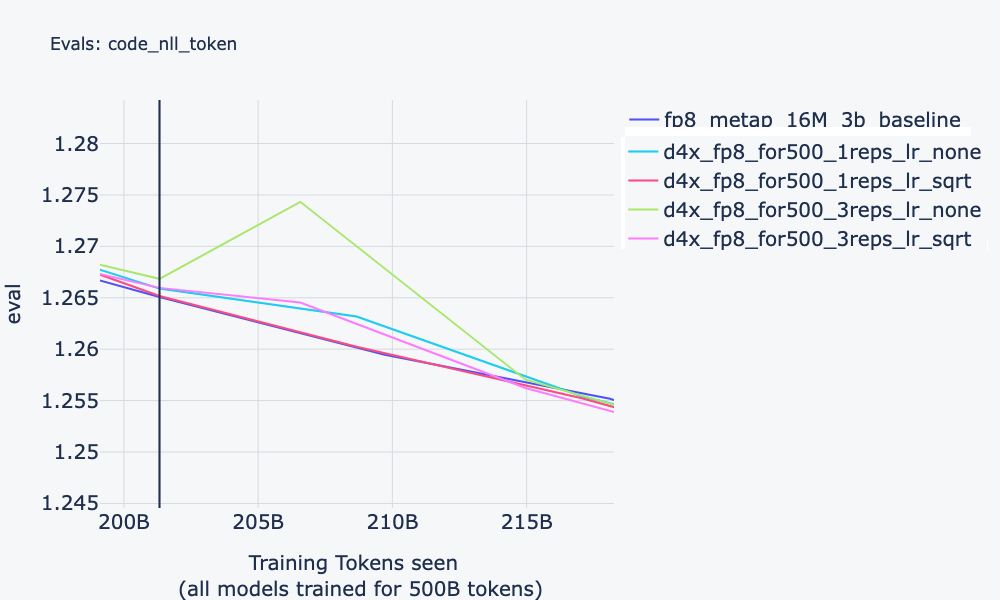}
     \caption{Training progress with short failures.}
     \label{fig:short-failure}
     \end{subfigure}
     ~
     \begin{subfigure}[b]{0.45\textwidth}
     \includegraphics[width=\textwidth]{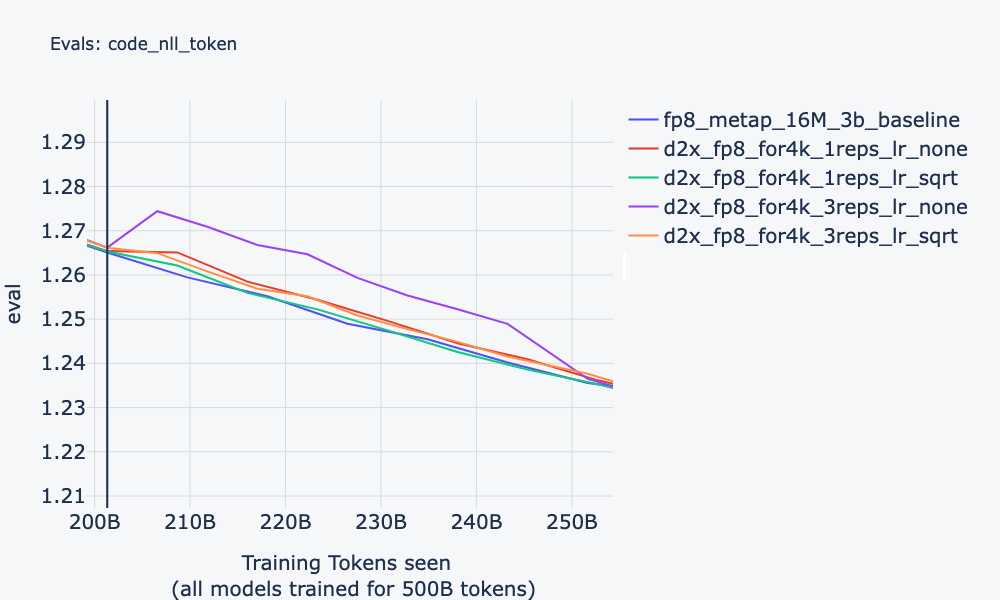}
     \caption{Training progress with long failures.}
     \label{fig:long-failure}
     \end{subfigure}
     \caption{Training progress fluctuation when zoomed in.}
     \label{fig:accuracy2}
     \vspace{-.1in}
\end{figure*}

Then we zoon into a small window of training progress for better observability,
as shown in Figure~\ref{fig:accuracy2}. Figure~\ref{fig:short-failure} shows that fluctuation
occurs in a short window, and three concurrent failures cause a greater fluctuation
than one failure, which is as expected. Here, we can also observe the effectiveness of
the learning rate intervention strategy: $sqrt$ helps to
flatten the curve, which might be helpful for the purpose of debugging.
Figure~\ref{fig:long-failure} shows that the conclusion still holds when we have
longer failures. Even in the $d2x\_fp8\_for4k\_3reps\_lr\_X$ settings with the most aggressive
failure pattern, the training accuracy eventually goes back to the baseline.

In summary, we believe that asynchronous recovery will not significantly degrade training
accuracy in the end, though it may cause some fluctuation during training. We recommend
square root learning rate intervention to flatten such fluctuation.
In other settings, such as those with fp16 or different evaluation sets, we arrive at the same conclusion.

\subsection{FTAR performance}
\label{sec:eval-ftar}

Since FTAR is a core component of \sys and is performance critical, we measure its performance
with a micro benchmark, which allocates GPUs in different AI zones and ask them to transfers messages with different sizes
using FTAR. 
As shown in Table~\ref{tab:ftar}, the performance of FTAR is close
to that of native NCCL implementation.

\begin{table}
\centering
\resizebox{0.48\textwidth}{!}{
\begin{tabular}{|c|c|c|c|c|c|c|c|c|}
\hline
            & \multicolumn{2}{|c|}{2 Ranks} & \multicolumn{2}{|c|}{4 Ranks} & \multicolumn{2}{|c|}{8 Ranks} & \multicolumn{2}{|c|}{16 Ranks}\\
\hline
           &   FTAR & NCCL &   FTAR & NCCL &   FTAR & NCCL &   FTAR & NCCL \\
\hline
256MB & 40.06 & 40.53 & 41.17 & 41.2 & 40.12 & 41.82 & 41.25 & 41.17 \\
\hline
512MB &  39.8 & 42.12 & 42.15 & 42.32 & 43.01 & 43.61 & 43.61 & 43.15 \\
\hline
1GB &  40.93 & 43.87 & 44.67 & 43.93 & 44.67 & 44.87 & 45.51 & 44.32 \\
\hline
\end{tabular}
}
\caption{FTAR bandwidth (GB/s). Average of 100 runs.}
\label{tab:ftar}
\end{table}

\section{Related work}\label{sec:related-work}

\vspace{-.1in}
\paragraph{Fault tolerant training.}
As discussed, many training systems use synchronous checkpoint based recovery~\cite{megascale,RobustLLM,unicorn,tpu-resiliency,bloom} and thus face a similar challenge to
our baseline system. Notably, ByteDance reported that it uses
the same design in a recent publication~\cite{RobustLLM}.

Some works have explored the idea elastic training, which naturally
enables asynchronous recovery~\cite{sergeev2018horovodfasteasydistributed,pytorch-distributed,shukla2022singularity}.
For example, Horovod~\cite{sergeev2018horovodfasteasydistributed} supports adding or removing nodes by leveraging data parallelism,
similar as \sys.
However, they require either a restart from the latest checkpoint or an expensive re-shuffle
of model states.
Furthermore, they have not considered the challenges of realizing this idea
at a large scale, such as the long latency of rebuidling NCCL links.
In production systems, Google Gemini reports that ``Our system now automatically continues training with fewer
slices of TPU chips when there is a localized failure''~\cite{comanici2025gemini25pushingfrontier}, but
does not disclose further details.
Oobleck and ReCycle leverage pipeline parallelism~\cite{Jang2023Oobleck,Gandhi2024ReCycle}
to achieve fault tolerance by rebuilding pipelines when failures occur.
However, it has not considered long networking latency and limited bandwidth across DCs, which may
make pipelines across DCs less efficient. Furthermore, rebuilding connections
would still be a challenge for NCCL based implementations.
Bamboo applies redundant computation to achieve fault tolerance with the
cost of performance in the failure-free case~\cite{Thorpe2023Bamboo}.

\vspace{-.1in}
\paragraph{All reduce algorithms and implementation.}
There has been an extensive study of all reduce algorithms in both
the High Performance Computing (HPC) field and the ML training field, including ring based algorithms
and other variations~\cite{Patarasuk2009Bandwidth,Thakur2005MPIOptimization,Cho2019BlueConnect,Jiang2020HRA,Ueno2019Study}.

In terms of GPU-related all reduce implementations, early systems use CPU-driven designs as GPUs in early days lack
capability to perform I/O~\cite{Wang2011MVAPICH2-GPU}. Therefore, the CPUs have to copy data from GPU memory to
main memory and perform I/Os. With the introduction of techniques like GPUDirect,
later systems have started to offload more work to GPUs~\cite{Potluri2013GPUDirect,Wang2014GPUAware,Shi2014SmallMessage}.
This trend finally leads to NCCL, which can almost completely bypass CPUs and become the de facto
communication library in large-scale training.
However, as shown in our work, GPU-driven implementation still
lacks the capability to implement complicated logic necessary for fault tolerance.
One key contribution of our work is to show that, with a hybrid design, which
relies on CPU for the control plane and GPU for the data plane, can achieve the
benefits of both---the capability to implement complicated control logic and a
performance comparable to native NCCL.

\vspace{-.1in}
\paragraph{Others.} \sys is related to work in many areas,
including parallel training~\cite{fsdp,shoeybi2019megatron,controlled_memory_zero_bubble_pp,deepseekai2025deepseekv3technicalreport,dao2023flashattention,shah2024flashattention}, efficient checkpoints~\cite{checkfreq,check-n-run,Wang2023AmazonGemini},
emulating a large-scale system with mocking~\cite{Wang2025GPEmu,dynamometer,Wang2014Exalt,Agrawal2011David,Stuardo2019ScaleCheck},
learning rate adjustment~\cite{smith2018dontdecaylearningrate,lin2019dynamicminibatchsgdelastic,goyal2018accuratelargeminibatchsgd,devarakonda2018adabatchadaptivebatchsizes}, root cause analysis of failures in training~\cite{Zhou2025Ascend,RobustLLM}, etc.
\sys leverage ideas from some of them.

\section{Conclusion}

This paper explores the challenges of scaling LLM training to 100K GPUs.
To address these challenges, \sys adopts the asynchronous recovery paradigm
and realizes this idea with a CPU-GPU hybrid all-reduce protocol,
a non-blocking catch-up protocol, and a number of other optimizations. Our evaluation
with 100K GPUs has demonstrated that such a design can significantly
reduce stalling time. Our experiments
at a smaller scale suggest that frequent asynchronous recovery does not
degrade training performance.


\bibliography{paper,references,ref2}

@article{shah2024flashattention,
  title={FlashAttention-3: Fast and accurate attention with asynchrony and low-precision},
  author={Shah, Jay and Bikshandi, Ganesh and Zhang, Ying and Thakkar, Vijay and Ramani, Pradeep and Dao, Tri},
  journal={Advances in Neural Information Processing Systems},
  volume={37},
  pages={68658--68685},
  year={2024}
}

@article{dao2023flashattention,
  title={FlashAttention-2: Faster attention with better parallelism and work partitioning},
  author={Dao, Tri},
  journal={arXiv preprint arXiv:2307.08691},
  year={2023}
}

@article{fsdp,
  author = {Zhao, Yanli and Gu, Andrew and Varma, Rohan and Luo, Liang and Huang, Chien-Chin and Xu, Min and Wright, Less and Shojanazeri, Hamid and Ott, Myle and Shleifer, Sam and Desmaison, Alban and Balioglu, Can and Damania, Pritam and Nguyen, Bernard and Chauhan, Geeta and Hao, Yuchen and Mathews, Ajit and Li, Shen},
  title = {PyTorch FSDP: Experiences on Scaling Fully Sharded Data Parallel},
  year = {2023},
  publisher = {VLDB Endowment},
  volume = {16},
  number = {12},
  journal = {Proceedings of the VLDB Endowment},
  pages = {3848–3860},
}

@misc{controlled_memory_zero_bubble_pp,
      title={Pipeline Parallelism with Controllable Memory}, 
      author={Penghui Qi and Xinyi Wan and Nyamdavaa Amar and Min Lin},
      year={2024},
      eprint={2405.15362},
      archivePrefix={arXiv},
      primaryClass={cs.LG},
      url={https://arxiv.org/abs/2405.15362}, 
}

@article{shoeybi2019megatron,
  title={Megatron-lm: Training multi-billion parameter language models using model parallelism},
  author={Shoeybi, Mohammad and Patwary, Mostofa and Puri, Raul and LeGresley, Patrick and Casper, Jared and Catanzaro, Bryan},
  journal={arXiv preprint arXiv:1909.08053},
  year={2019}
}

@article{lin2019dynamicminibatchsgdelastic,
  title={Dynamic Mini-batch SGD for Elastic Distributed Training: Learning in the Limbo of Resources
},
  author={Lin, Haibin and Zhang, Hang and Ma, Yifei and He, Tong and Zhang, Zhi and Zha, Sheng and Li, Mu},
  journal={arXiv preprint arXiv:1904.12043},
  year={2019}
}

@inproceedings{gangidiRDMA,
author = {Gangidi, Adithya and Miao, Rui and Zheng, Shengbao and Bondu, Sai Jayesh and Goes, Guilherme and Morsy, Hany and Puri, Rohit and Riftadi, Mohammad and Shetty, Ashmitha Jeevaraj and Yang, Jingyi and Zhang, Shuqiang and Fernandez, Mikel Jimenez and Gandham, Shashidhar and Zeng, Hongyi},
title = {RDMA over Ethernet for Distributed Training at Meta Scale},
year = {2024},
isbn = {9798400706141},
publisher = {Association for Computing Machinery},
address = {New York, NY, USA},
url = {https://doi.org/10.1145/3651890.3672233},
doi = {10.1145/3651890.3672233},
booktitle = {Proceedings of the ACM SIGCOMM 2024 Conference},
pages = {57–70},
numpages = {14},
location = {Sydney, NSW, Australia},
series = {ACM SIGCOMM '24}
}

@inproceedings{deng2025minder,
  title={Minder: Faulty machine detection for large-scale distributed model training},
  author={Deng, Yangtao and Shi, Xiang and Jiang, Zhuo and Zhang, Xingjian and Zhang, Lei and Zhang, Zhang and Li, Bo and Song, Zuquan and Zhu, Hang and Liu, Gaohong and others},
  booktitle={22nd USENIX Symposium on Networked Systems Design and Implementation (NSDI 25)},
  pages={505--521},
  year={2025}
}

@article{llama3,
  title={{The Llama 3 Herd of Models}},
  author={{Llama Team, AI @ Meta}},
  journal={arXiv preprint arXiv:2407.21783},
  year={2024}
}

@inproceedings{llama3-parallelism,
title = {{Scaling Llama 3 Training with Efficient Parallelism Strategies}},
author = {Weiwei Chu and Xinfeng Xie and Jiecao Yu and Jie Wang and Amar Phanishayee and Chunqiang Tang and Yuchen Hao and Jianyu Huang and Mustafa Ozdal and Jun Wang and Vedanuj Goswami and Naman Goyal and Abhishek Kadian and Andrew Gu and Chris Cai and Feng Tian and Xiaodong Wang and Min Si and Pavan
Balaji and Ching-Hsiang Chu and Jongsoo Park},
booktitle = {proceedings of the 2025 International Symposium on Computer Architecture},
year = 2025
}

@String{Computing = "Computing" }

@String{Computer = "{IEEE} Computer" }

@String{Academic = "Academic Press" }

@inproceedings {megascale,
author = {Ziheng Jiang and Haibin Lin and Yinmin Zhong and Qi Huang and Yangrui Chen and Zhi Zhang and Yanghua Peng and Xiang Li and Cong Xie and Shibiao Nong and Yulu Jia and Sun He and Hongmin Chen and Zhihao Bai and Qi Hou and Shipeng Yan and Ding Zhou and Yiyao Sheng and Zhuo Jiang and Haohan Xu and Haoran Wei and Zhang Zhang and Pengfei Nie and Leqi Zou and Sida Zhao and Liang Xiang and Zherui Liu and Zhe Li and Xiaoying Jia and Jianxi Ye and Xin Jin and Xin Liu},
title = {{MegaScale}: Scaling Large Language Model Training to More Than 10,000 {GPUs}},
booktitle = {21st USENIX Symposium on Networked Systems Design and Implementation (NSDI 24)},
year = {2024},
isbn = {978-1-939133-39-7},
address = {Santa Clara, CA},
pages = {745--760},
url = {https://www.usenix.org/conference/nsdi24/presentation/jiang-ziheng},
publisher = {USENIX Association},
month = apr
}

@inproceedings {check-n-run,
author = {Assaf Eisenman and Kiran Kumar Matam and Steven Ingram and Dheevatsa Mudigere and Raghuraman Krishnamoorthi and Krishnakumar Nair and Misha Smelyanskiy and Murali Annavaram},
title = {{Check-N-Run}: a Checkpointing System for Training Deep Learning Recommendation Models},
booktitle = {19th USENIX Symposium on Networked Systems Design and Implementation (NSDI 22)},
year = {2022},
isbn = {978-1-939133-27-4},
address = {Renton, WA},
pages = {929--943},
url = {https://www.usenix.org/conference/nsdi22/presentation/eisenman},
publisher = {USENIX Association},
month = apr
}

@inproceedings {checkfreq,
author = {Jayashree Mohan and Amar Phanishayee and Vijay Chidambaram},
title = {{CheckFreq}: Frequent, {Fine-Grained} {DNN} Checkpointing},
booktitle = {19th USENIX Conference on File and Storage Technologies (FAST 21)},
year = {2021},
isbn = {978-1-939133-20-5},
pages = {203--216},
url = {https://www.usenix.org/conference/fast21/presentation/mohan},
publisher = {USENIX Association},
month = feb
}

@ARTICLE{unicorn,
       author = {{He}, Tao and {Li}, Xue and {Wang}, Zhibin and {Qian}, Kun and {Xu}, Jingbo and {Yu}, Wenyuan and {Zhou}, Jingren},
        title = "{Unicron: Economizing Self-Healing LLM Training at Scale}",
      journal = {arXiv e-prints},
         year = 2023,
        month = dec,
          eid = {arXiv:2401.00134},
        pages = {arXiv:2401.00134},
          doi = {10.48550/arXiv.2401.00134},
archivePrefix = {arXiv},
       eprint = {2401.00134},
 primaryClass = {cs.DC},
       adsurl = {https://ui.adsabs.harvard.edu/abs/2024arXiv240100134H}
}

@article{pytorch-distributed,
author = {Li, Shen and Zhao, Yanli and Varma, Rohan and Salpekar, Omkar and Noordhuis, Pieter and Li, Teng and Paszke, Adam and Smith, Jeff and Vaughan, Brian and Damania, Pritam and Chintala, Soumith},
title = {PyTorch distributed: experiences on accelerating data parallel training},
year = {2020},
issue_date = {August 2020},
publisher = {VLDB Endowment},
volume = {13},
number = {12},
issn = {2150-8097},
url = {https://doi.org/10.14778/3415478.3415530},
doi = {10.14778/3415478.3415530},
journal = {Proc. VLDB Endow.},
month = {aug},
pages = {3005–3018},
numpages = {14}
}

@ARTICLE{bloom,
       author = { {Le Scao}, Teven and others},
        title = "{BLOOM: A 176B-Parameter Open-Access Multilingual Language Model}",
      journal = {arXiv e-prints},
         year = 2022,
        month = nov,
          eid = {arXiv:2211.05100},
        pages = {arXiv:2211.05100},
          doi = {10.48550/arXiv.2211.05100},
archivePrefix = {arXiv},
       eprint = {2211.05100},
 primaryClass = {cs.CL},
       adsurl = {https://ui.adsabs.harvard.edu/abs/2022arXiv221105100W}
}

@inproceedings {tpu-resiliency,
author = {Yazhou Zu and Alireza Ghaffarkhah and Hoang-Vu Dang and Brian Towles and Steven Hand and Safeen Huda and Adekunle Bello and Alexander Kolbasov and Arash Rezaei and Dayou Du and Steve Lacy and Hang Wang and Aaron Wisner and Chris Lewis and Henri Bahini},
title = {Resiliency at Scale: Managing {Google{\textquoteright}s} {TPUv4} Machine Learning Supercomputer},
booktitle = {21st USENIX Symposium on Networked Systems Design and Implementation (NSDI 24)},
year = {2024},
isbn = {978-1-939133-39-7},
address = {Santa Clara, CA},
pages = {761--774},
url = {https://www.usenix.org/conference/nsdi24/presentation/zu},
publisher = {USENIX Association},
month = apr
}

@inproceedings{RobustLLM,
author = {Wan, Borui and Liu, Gaohong and Song, Zuquan and Wang, Jun and Zhang, Yun and Sheng, Guangming and Wang, Shuguang and Wei, Houmin and Wang, Chenyuan and Lou, Weiqiang and Yang, Xi and Zhang, Mofan and Jiang, Kaihua and Ren, Cheng and Zhi, Xiaoyun and Yu, Menghan and Nan, Zhe and Zheng, Zhuolin and Zhong, Baoquan and Wang, Qinlong and Yu, Huan and Chi, Jinxin and Zhang, Wang and Li, Yuhan and Du, Zixian and Zhao, Sida and Zhang, Yongqiang and Tang, Jingzhe and Liu, Zherui and Wu, Chuan and Peng, Yanghua and Lin, Haibin and Xiao, Wencong and Liu, Xin and Xiang, Liang},
title = {{Robust LLM Training Infrastructure at ByteDance}},
year = {2025},
isbn = {9798400718700},
publisher = {Association for Computing Machinery},
address = {New York, NY, USA},
url = {https://doi.org/10.1145/3731569.3764838},
doi = {10.1145/3731569.3764838},
booktitle = {Proceedings of the ACM SIGOPS 31st Symposium on Operating Systems Principles},
pages = {186–203},
numpages = {18},
location = {Lotte Hotel World, Seoul, Republic of Korea},
series = {SOSP '25}
}

@misc{deepseekai2025deepseekv3technicalreport,
      title={DeepSeek-V3 Technical Report}, 
      author={{Aixin Liu, et. al}},
      year={2025},
      eprint={2412.19437},
      archivePrefix={arXiv},
      primaryClass={cs.CL},
      url={https://arxiv.org/abs/2412.19437}, 
}

@inproceedings{Jang2023Oobleck,
author = {Jang, Insu and Yang, Zhenning and Zhang, Zhen and Jin, Xin and Chowdhury, Mosharaf},
title = {{Oobleck: Resilient Distributed Training of Large Models Using Pipeline Templates}},
year = {2023},
isbn = {9798400702297},
publisher = {Association for Computing Machinery},
address = {New York, NY, USA},
url = {https://doi.org/10.1145/3600006.3613152},
doi = {10.1145/3600006.3613152},
booktitle = {Proceedings of the 29th Symposium on Operating Systems Principles},
pages = {382–395},
numpages = {14},
location = {Koblenz, Germany},
series = {SOSP '23}
}

@inproceedings{Gandhi2024ReCycle,
author = {Gandhi, Swapnil and Zhao, Mark and Skiadopoulos, Athinagoras and Kozyrakis, Christos},
title = {{ReCycle: Resilient Training of Large DNNs using Pipeline Adaptation}},
year = {2024},
isbn = {9798400712517},
publisher = {Association for Computing Machinery},
address = {New York, NY, USA},
url = {https://doi.org/10.1145/3694715.3695960},
doi = {10.1145/3694715.3695960},
booktitle = {Proceedings of the ACM SIGOPS 30th Symposium on Operating Systems Principles},
pages = {211–228},
numpages = {18},
location = {Austin, TX, USA},
series = {SOSP '24}
}

@article{Thakur2005MPIOptimization,
author = {Thakur, Rajeev and Rabenseifner, Rolf and Gropp, William},
title = {Optimization of Collective Communication Operations in MPICH},
year = {2005},
issue_date = {February 2005},
publisher = {Sage Publications, Inc.},
address = {USA},
volume = {19},
number = {1},
issn = {1094-3420},
url = {https://doi.org/10.1177/1094342005051521},
doi = {10.1177/1094342005051521},
journal = {Int. J. High Perform. Comput. Appl.},
month = feb,
pages = {49–66},
numpages = {18}
}

@article{Patarasuk2009Bandwidth,
author = {Patarasuk, Pitch and Yuan, Xin},
title = {Bandwidth optimal all-reduce algorithms for clusters of workstations},
year = {2009},
issue_date = {February, 2009},
publisher = {Academic Press, Inc.},
address = {USA},
volume = {69},
number = {2},
issn = {0743-7315},
url = {https://doi.org/10.1016/j.jpdc.2008.09.002},
doi = {10.1016/j.jpdc.2008.09.002},
journal = {J. Parallel Distrib. Comput.},
month = feb,
pages = {117–124},
numpages = {8}
}

@inproceedings {Thorpe2023Bamboo,
author = {John Thorpe and Pengzhan Zhao and Jonathan Eyolfson and Yifan Qiao and Zhihao Jia and Minjia Zhang and Ravi Netravali and Guoqing Harry Xu},
title = {Bamboo: Making Preemptible Instances Resilient for Affordable Training of Large {DNNs}},
booktitle = {20th USENIX Symposium on Networked Systems Design and Implementation (NSDI 23)},
year = {2023},
isbn = {978-1-939133-33-5},
address = {Boston, MA},
pages = {497--513},
url = {https://www.usenix.org/conference/nsdi23/presentation/thorpe},
publisher = {USENIX Association},
month = apr
}

@misc{comanici2025gemini25pushingfrontier,
      title={Gemini 2.5: Pushing the Frontier with Advanced Reasoning, Multimodality, Long Context, and Next Generation Agentic Capabilities}, 
      author={{Gheorghe Comanici et. al.}},
      year={2025},
      eprint={2507.06261},
      archivePrefix={arXiv},
      primaryClass={cs.CL},
      url={https://arxiv.org/abs/2507.06261}, 
}

@article{Wang2011MVAPICH2-GPU,
author = {Wang, Hao and Potluri, Sreeram and Luo, Miao and Singh, Ashish Kumar and Sur, Sayantan and Panda, Dhabaleswar K.},
title = {MVAPICH2-GPU: optimized GPU to GPU communication for InfiniBand clusters},
year = {2011},
issue_date = {June 2011},
publisher = {Springer-Verlag},
address = {Berlin, Heidelberg},
volume = {26},
number = {3–4},
issn = {1865-2034},
url = {https://doi.org/10.1007/s00450-011-0171-3},
doi = {10.1007/s00450-011-0171-3},
journal = {Comput. Sci.},
month = jun,
pages = {257–266},
numpages = {10}
}

@misc{llama4,
  author = {Meta},
  title = {{The Llama 4 herd: The beginning of a new era of natively multimodal AI innovation}},
  year = {2025},
  url={https://ai.meta.com/blog/llama-4-multimodal-intelligence/}
}

@misc{semianalysis,
  author = {Dylan Patel and Daniel Nishball},
  title = {{100,000 H100 Clusters: Power, Network Topology, Ethernet vs InfiniBand, Reliability, Failures, Checkpointing}},
  year = {2024},
  url={https://newsletter.semianalysis.com/p/100000-h100-clusters-power-network}
}

@misc{200kGPUs,
  author = {{AI News Hub}},
  title = {{Grok 4 and Colossus 2: xAI’s Groundbreaking Gigawatt AI Training Supercluster Unveiled for 2025}},
  year = {2025},
  url={https://www.ainewshub.org/post/grok-4-and-colossus-2-xai-s-groundbreaking-gigawatt-ai-training-supercluster-unveiled-for-2025}
}

@misc{hsdp-aws,
  author = {{Amazon Web Services}},
  title = {{Hybrid sharded data parallelism}},
  year = {2025},
  url={https://docs.aws.amazon.com/sagemaker/latest/dg/model-parallel-core-features-v2-sharded-data-parallelism.html}
}

@misc{sergeev2018horovodfasteasydistributed,
      title={Horovod: fast and easy distributed deep learning in TensorFlow}, 
      author={Alexander Sergeev and Mike Del Balso},
      year={2018},
      eprint={1802.05799},
      archivePrefix={arXiv},
      primaryClass={cs.LG},
      url={https://arxiv.org/abs/1802.05799}, 
}

@inproceedings {Wang2012Gnothi,
author = {Yang Wang and Lorenzo Alvisi and Mike Dahlin},
title = {Gnothi: Separating Data and Metadata for Efficient and Available Storage Replication},
booktitle = {2012 USENIX Annual Technical Conference (USENIX ATC 12)},
year = {2012},
isbn = {978-931971-93-5},
address = {Boston, MA},
pages = {413--424},
url = {https://www.usenix.org/conference/atc12/technical-sessions/presentation/wang},
publisher = {USENIX Association},
month = jun
}

@inproceedings {Shi2016Cheap,
author = {Rong Shi and Yang Wang},
title = {Cheap and Available State Machine Replication},
booktitle = {2016 USENIX Annual Technical Conference (USENIX ATC 16)},
year = {2016},
isbn = {978-1-931971-30-0},
address = {Denver, CO},
pages = {265--279},
url = {https://www.usenix.org/conference/atc16/technical-sessions/presentation/shi},
publisher = {USENIX Association},
month = jun
}

@inproceedings{Wang2023SDC,
author = {Wang, Shaobu and Zhang, Guangyan and Wei, Junyu and Wang, Yang and Wu, Jiesheng and Luo, Qingchao},
title = {Understanding Silent Data Corruptions in a Large Production CPU Population},
year = {2023},
isbn = {9798400702297},
publisher = {Association for Computing Machinery},
address = {New York, NY, USA},
url = {https://doi.org/10.1145/3600006.3613149},
doi = {10.1145/3600006.3613149},
booktitle = {Proceedings of the 29th Symposium on Operating Systems Principles},
pages = {216–230},
numpages = {15},
location = {Koblenz, Germany},
series = {SOSP '23}
}

@inproceedings{Burrow2006Chubby,
author = {Burrows, Mike},
title = {The Chubby lock service for loosely-coupled distributed systems},
year = {2006},
isbn = {1931971471},
publisher = {USENIX Association},
address = {USA},
booktitle = {Proceedings of the 7th Symposium on Operating Systems Design and Implementation},
pages = {335–350},
numpages = {16},
location = {Seattle, Washington},
series = {OSDI '06}
}

@inproceedings {Balakrishnan2020Delos,
author = {Mahesh Balakrishnan and Jason Flinn and Chen Shen and Mihir Dharamshi and Ahmed Jafri and Xiao Shi and Santosh Ghosh and Hazem Hassan and Aaryaman Sagar and Rhed Shi and Jingming Liu and Filip Gruszczynski and Xianan Zhang and Huy Hoang and Ahmed Yossef and Francois Richard and Yee Jiun Song},
title = {Virtual Consensus in Delos},
booktitle = {14th USENIX Symposium on Operating Systems Design and Implementation (OSDI 20)},
year = {2020},
isbn = {978-1-939133-19-9},
pages = {617--632},
url = {https://www.usenix.org/conference/osdi20/presentation/balakrishnan},
publisher = {USENIX Association},
month = nov
}

@inproceedings{Wang2014Exalt,
author = {Wang, Yang and Kapritsos, Manos and Schmidt, Lara and Alvisi, Lorenzo and Dahlin, Mike},
title = {Exalt: empowering researchers to evaluate large-scale storage systems},
year = {2014},
isbn = {9781931971096},
publisher = {USENIX Association},
address = {USA},
booktitle = {Proceedings of the 11th USENIX Conference on Networked Systems Design and Implementation},
pages = {129–141},
numpages = {13},
location = {Seattle, WA},
series = {NSDI'14}
}

@inproceedings{Agrawal2011David,
author = {Agrawal, Nitin and Arulraj, Leo and Arpaci-Dusseau, Andrea C. and Arpaci-Dusseau, Remzi H.},
title = {Emulating Goliath storage systems with David},
year = {2011},
isbn = {9781931971829},
publisher = {USENIX Association},
address = {USA},
booktitle = {Proceedings of the 9th USENIX Conference on File and Stroage Technologies},
pages = {15},
numpages = {1},
location = {San Jose, California},
series = {FAST'11}
}

@inproceedings {Stuardo2019ScaleCheck,
author = {Cesar A. Stuardo and Tanakorn Leesatapornwongsa and Riza O. Suminto and Huan Ke and Jeffrey F. Lukman and Wei-Chiu Chuang and Shan Lu and Haryadi S. Gunawi},
title = {{ScaleCheck}: A {Single-Machine} Approach for Discovering Scalability Bugs in Large Distributed Systems},
booktitle = {17th USENIX Conference on File and Storage Technologies (FAST 19)},
year = {2019},
isbn = {978-1-939133-09-0},
address = {Boston, MA},
pages = {359--373},
url = {https://www.usenix.org/conference/fast19/presentation/stuardo},
publisher = {USENIX Association},
month = feb
}

@misc{dynamometer,
      title={{Dynamometer: Scale Testing HDFS on Minimal Hardware with Maximum Fidelity}}, 
      author={Erik Krogen},
      year={2018},
      url={https://www.linkedin.com/blog/engineering/archive/dynamometer-scale-testing-hdfs-on-minimal-hardware-with-maximum}, 
}

@inproceedings{Wang2023AmazonGemini,
author = {Wang, Zhuang and Jia, Zhen and Zheng, Shuai and Zhang, Zhen and Fu, Xinwei and Ng, T. S. Eugene and Wang, Yida},
title = {GEMINI: Fast Failure Recovery in Distributed Training with In-Memory Checkpoints},
year = {2023},
isbn = {9798400702297},
publisher = {Association for Computing Machinery},
address = {New York, NY, USA},
url = {https://doi.org/10.1145/3600006.3613145},
doi = {10.1145/3600006.3613145},
booktitle = {Proceedings of the 29th Symposium on Operating Systems Principles},
pages = {364–381},
numpages = {18},
location = {Koblenz, Germany},
series = {SOSP '23}
}

@article{shukla2022singularity,
      title={{Singularity: Planet-Scale, Preemptive and Elastic Scheduling of AI Workloads}},
      author={Dharma Shukla and Muthian Sivathanu and Srinidhi Viswanatha and Bhargav Gulavani and Rimma Nehme and Amey Agrawal and Chen Chen and Nipun Kwatra and Ramachandran Ramjee and Pankaj Sharma and Atul Katiyar and Vipul Modi and Vaibhav Sharma and Abhishek Singh and Shreshth Singhal and Kaustubh Welankar and Lu Xun and Ravi Anupindi and Karthik Elangovan and Hasibur Rahman and Zhou Lin and Rahul Seetharaman and Cheng Xu and Eddie Ailijiang and Suresh Krishnappa and Mark Russinovich},
      year={2022},
      eprint={2202.07848},
      archivePrefix={arXiv},
      primaryClass={cs.DC},
      journal={arXiv preprint arXiv:2202.07848}
}

@inproceedings{Zhou2025Ascend,
author = {Zhou, Yuhang and Wang, Zibo and Wang, Zhibin and Zhang, Ruyi and Tian, Chen and Wang, Xiaoliang and Dou, Wanchun and Chen, Guihai and Wang, Bingqiang and Tian, Yonghong and Zhang, Yan and Wang, Hui and Wei, Fuchun and Sun, Boquan and Zhang, Jingyi and She, Bin and Su, Teng and Yao, Yifan and Li, Chunsheng and Zhang, Ziyang and Wang, Yaoyuan and Zhou, Bin and Liu, Guyue},
title = {{Accelerating model training on ascend chips: an industrial system for profiling, analysis and optimization}},
year = {2025},
isbn = {978-1-939133-48-9},
publisher = {USENIX Association},
address = {USA},
booktitle = {Proceedings of the 2025 USENIX Conference on Usenix Annual Technical Conference},
articleno = {82},
numpages = {22},
location = {Boston, MA, USA},
series = {USENIX ATC '25}
}

@misc{smith2018dontdecaylearningrate,
      title={{Don't Decay the Learning Rate, Increase the Batch Size}}, 
      author={Samuel L. Smith and Pieter-Jan Kindermans and Chris Ying and Quoc V. Le},
      year={2018},
      eprint={1711.00489},
      archivePrefix={arXiv},
      primaryClass={cs.LG},
      url={https://arxiv.org/abs/1711.00489}, 
}

@misc{goyal2018accuratelargeminibatchsgd,
      title={{Accurate, Large Minibatch SGD: Training ImageNet in 1 Hour}}, 
      author={Priya Goyal and Piotr Dollár and Ross Girshick and Pieter Noordhuis and Lukasz Wesolowski and Aapo Kyrola and Andrew Tulloch and Yangqing Jia and Kaiming He},
      year={2018},
      eprint={1706.02677},
      archivePrefix={arXiv},
      primaryClass={cs.CV},
      url={https://arxiv.org/abs/1706.02677}, 
}

@misc{devarakonda2018adabatchadaptivebatchsizes,
      title={{AdaBatch: Adaptive Batch Sizes for Training Deep Neural Networks}}, 
      author={Aditya Devarakonda and Maxim Naumov and Michael Garland},
      year={2018},
      eprint={1712.02029},
      archivePrefix={arXiv},
      primaryClass={cs.LG},
      url={https://arxiv.org/abs/1712.02029}, 
}

@inproceedings{Cho2019BlueConnect,
 author = {Cho, Minsik and Finkler, Ulrich and Kung, David and Hunter, Hillery},
 booktitle = {Proceedings of Machine Learning and Systems},
 editor = {A. Talwalkar and V. Smith and M. Zaharia},
 pages = {241--251},
 title = {{BlueConnect: Decomposing All-Reduce for Deep Learning on Heterogeneous Network Hierarchy}},
 url = {https://proceedings.mlsys.org/paper_files/paper/2019/file/0c8abcf158ed12d0dd94480681186fda-Paper.pdf},
 volume = {1},
 year = {2019}
}

@ARTICLE{Jiang2020HRA,
  author={Jiang, Youhe and Gu, Huaxi and Lu, Yunfeng and Yu, Xiaoshan},
  journal={IEEE Access}, 
  title={2D-HRA: Two-Dimensional Hierarchical Ring-Based All-Reduce Algorithm in Large-Scale Distributed Machine Learning}, 
  year={2020},
  volume={8},
  number={},
  pages={183488-183494},
  doi={10.1109/ACCESS.2020.3028367}}

@INPROCEEDINGS{Ueno2019Study,
  author={Ueno, Yuichiro and Yokota, Rio},
  booktitle={2019 19th IEEE/ACM International Symposium on Cluster, Cloud and Grid Computing (CCGRID)}, 
  title={Exhaustive Study of Hierarchical AllReduce Patterns for Large Messages Between GPUs}, 
  year={2019},
  volume={},
  number={},
  pages={430-439},
  doi={10.1109/CCGRID.2019.00057}}

@INPROCEEDINGS{Potluri2013GPUDirect,
  author={Potluri, Sreeram and Hamidouche, Khaled and Venkatesh, Akshay and Bureddy, Devendar and Panda, Dhabaleswar K.},
  booktitle={2013 42nd International Conference on Parallel Processing}, 
  title={Efficient Inter-node MPI Communication Using GPUDirect RDMA for InfiniBand Clusters with NVIDIA GPUs}, 
  year={2013},
  volume={},
  number={},
  pages={80-89},
  doi={10.1109/ICPP.2013.17}}

@ARTICLE{Wang2014GPUAware,
  author={Wang, Hao and Potluri, Sreeram and Bureddy, Devendar and Rosales, Carlos and Panda, Dhabaleswar K.},
  journal={IEEE Transactions on Parallel and Distributed Systems}, 
  title={GPU-Aware MPI on RDMA-Enabled Clusters: Design, Implementation and Evaluation}, 
  year={2014},
  volume={25},
  number={10},
  pages={2595-2605},
  doi={10.1109/TPDS.2013.222}}

@INPROCEEDINGS{Shi2014SmallMessage,
  author={Shi, Rong and Potluri, Sreeram and Hamidouche, Khaled and Perkins, Jonathan and Li, Mingzhe and Rossetti, Davide and Panda, Dhabaleswar K. D K},
  booktitle={2014 21st International Conference on High Performance Computing (HiPC)}, 
  title={Designing efficient small message transfer mechanism for inter-node MPI communication on InfiniBand GPU clusters}, 
  year={2014},
  volume={},
  number={},
  pages={1-10},
  doi={10.1109/HiPC.2014.7116873}}

@misc{nccl,
  author = {NVIDIA},
  title = {NVIDIA Collective Communications Library (NCCL)},
  year = {2025},
  url={https://developer.nvidia.com/nccl}
}

@article{Wang2025GPEmu,
author = {Wang, Meng and Waldspurger, Gus and Ananda, Naufal and Huang, Yuyang and Wiharja, Kemas and Bent, John and Sundararaman, Swaminathan and Chidambaram, Vijay and Gunawi, Haryadi S.},
title = {GPEmu: A GPU Emulator for Faster and Cheaper Prototyping and Evaluation of Deep Learning System Research},
year = {2025},
issue_date = {February 2025},
publisher = {VLDB Endowment},
volume = {18},
number = {6},
issn = {2150-8097},
url = {https://doi.org/10.14778/3725688.3725716},
doi = {10.14778/3725688.3725716},
journal = {Proc. VLDB Endow.},
month = feb,
pages = {1919–1932},
numpages = {14}
}

\end{document}